\documentclass[aoas,preprint]{imsart}

\usepackage{amsthm,amsmath,natbib,graphicx}
\RequirePackage{natbib}
\usepackage{multirow}
\usepackage{multicol}
\usepackage{diagbox}
\usepackage{blkarray}
\usepackage{makecell}
\usepackage{vwcol}  

\usepackage{tikz}
\usetikzlibrary{bayesnet}

\startlocaldefs
\endlocaldefs

\begin{document}

\begin{frontmatter}

\title{Clustering Network Tree Data From Respondent-Driven Sampling With Application to Opioid Users in New York City}
\runtitle{Clustering Network Tree Data From Respondent-Driven Sampling With Application to Opioid Users in New York City}

\author{\fnms{Shuaimin} \snm{Kang}\ead[label=e1]{kang@math.umass.edu}}
\address{\printead{e1}}
\and 
\author{\fnms{Krista} \snm{Gile}\ead[label=e2]{gile@math.umass.edu}}
\address{\printead{e2}}
\and
\author{\fnms{Pedro} \snm{Mateu-Gelabert}\ead[label=e3]{mateu-gelabert@ndri.org}}
\address{\printead{e3}}
\and
\author{\fnms{Honoria} \snm{Guarino}\ead[label=e4]{guarino@ndri.org}}
\address{\printead{e4}}

\runauthor{}

\begin{abstract}
There is great interest in finding meaningful subgroups of attributed network data. There are many available methods for clustering complete network. Unfortunately, much network data is collected through sampling, and therefore incomplete. Respondent-driven sampling (RDS) is a widely used method for sampling hard-to-reach human populations based on tracing links in the underlying unobserved social network. The resulting data therefore have tree structure representing a sub-sample of the network, along with many nodal attributes. In this paper, we introduce an approach to adjust mixture models for general network clustering for samplings by RDS. We apply our model to data on opioid users in New York City, and detect communities reflecting group characteristics of interest for intervention activities, including drug use patterns, social connections and other community variables.
\end{abstract}

\end{frontmatter}

\section{Introduction}
Network clustering is used to detect groups within a graph where nodes in the same group have stronger social connections than nodes in different groups and where nodal attributes are more similar within groups. However, there are no existing methods for clustering social networks sampled with link-tracing mechanisms, such as Respondent-driven sampling (RDS). Traditional network clustering methods are not appropriate for RDS networks because of the link tracing procedure in RDS. Clustering of networks with node or edge features is well studied \citep{Yang2013}, \citep{Xu2012}, \citep{Qi2012}. In this paper, we build a mixture model for RDS network sample with node features, and add sampling weights to the likelihood to find clusters for the RDS network sample.\\
Respondent-driven sampling (RDS) \citep{Heckathorn1997} is a link-tracing network sampling method popularly used in sampling data from hard-to-reach populations, such as drug users and sex workers. It starts by selecting several people in the target population as seeds, then those seeds expand the sample by distributing coupons to people they know, those newly added samples distribute coupons in a similar way, and this process continues until reaching the desired sample size. Each coupon has a unique number which makes clear who recruited whom. RDS is a sampling method without replacement and its resulting observed network has tree structure with each tree starting with a different seed. The maximum number of coupons one person can distribute or the maximum number of people each person can recruit is usually small, like 3, to make sure the tree is deep enough, which helps reduce dependency of samples in a tree on its seed.\\
Each sampled person in the RDS network completes a survey, creating a node-attributed RDS network. Some node-attributed RDS networks have obvious homophily \citep{Gile2010}, which is the correlation between trait values of nodes connected by an edge. For example, in the opioid drug user RDS network, heavy drug users are more likely to be tied to, and therefore recruit heavy drug users.\\
Network clustering methods have been developed extensively. Maximizing modularity \citep{Newman2006}, minimizing cut \citep{Ding2001}, eigenvector related spectral clustering \citep{Ng2001} \citep{Shi2002}, and hierarchical clustering \citep{Bandyopadhyay2003} are widely used in computer science and biology to cluster complex graphs. Methods for clustering networks statistically through assigning distributions to network structures are also well developed. In the stochastic block model \citep{Nowicki1997}\citep{Karrer2011} \citep{Airoldi2008}, mixture and Bayesian mixture models \citep{Daudin2008}, edges follow Bernoulli or Bernoulli mixture distributions with the same connection probabilities if they're in the same block or community. Model based network clustering methods have also been used to cluster graphs with node or edge features. Handcock et al. (2007) models node pair connection probability as a logistic regression on covariates and the distance of the node pair in a latent social space. In Communities from Edge Structure and Node Attributes (CESNA) \citep{Yang2013}, links of the network and node attributes are modeled separately but connected by the node community membership probabilities. Xu et al. (2012) proposed a Bayesian probability model assuming network structure and node attributes are independent given node group status. In this paper, we build on Xu et al. (2012)'s assumption that node features and network structures are independent given node clustering status and build a mixture model from it. Since RDS generates incomplete network data with nodes and edges unequally sampled from a full network, the above network clustering methods are not valid. Therefore, we propose a weighted log-likelihood approach, adding nodal and edge inverse sampling probability weights (IPW) to the log-likelihood for inference. \\
In this paper, we are not only interested in clustering the RDS sample data, but also interested in the interpretation of those clusters and individuals within those clusters. To better interpret populations in each cluster, we should find and use less biased parameters given the sampled data. Weighting is a common way to reduce bias in sampled data. Weighted likelihood has been used in mixture models for reducing bias when outliers exist in the data \citep{Markatou2000}. The inverse selection probability-weighted likelihood method has also been studied for fitting sampled data \citep{Li2008} \citep{saegusa2013}. Weighted likelihood has been used for automatic model selection in density mixture clustering \citep{Cheung2005}. Weighted iterative clustering algorithms have also been well studied for better clustering \citep{Topchy2004}\citep{Zhang2001}\citep{Hamerly2002}. Based on those literatures and considering the un-equal sampling probabilities in RDS, the instances or nodes and edges in the RDS sample should not be treated equally. Therefore, we propose to add inverse sampling probabilities to the likelihood of the mixture model from the node attributed RDS sample data to approximate the likelihood in the pseduo-population, thus getting less biased parameter estimation and reasonable clustering. \\
In this paper, we review sampling probabilities in RDS in Section \ref{Background}. We propose a mixture model without weights as Benchmark model and extend the Benchmark model by adding IPW in Section \ref{Model}. Furthermore, we propose the weighted likelihood mixture model with tuning parameter to balance contribution of node features and network structure. In Section \ref{Evaluation}, we talk about evaluation of clustering algorithms and tuning parameter selection. In Section \ref{Simulation}, we compare the approaches proposed in Section \ref{Model} through simulation studies. In Section \ref{Application}, we apply our approach to opioid users' RDS data from New York City. In Section \ref{Conclusion} , we summarize the weighted log-likelihood mixture model for clustering incomplete node attributed RDS network data.
\section{Background}\label{Background}
\subsection{RDS network Structure and Notation}
As a link-tracing without replacement sampling method, RDS results in tree structured graphs as in the RDS network sample in Figure \ref{fig: figure 1}. Each person in the network is called a node. If two nodes are connected, we say there is an edge or a tie connecting them. In general, an adjacency matrix is used to describe connections between nodes in the network. Assume there are N nodes in the full network and $n (n\leq N)$ nodes in the RDS sample. Denote $Y = [y_{ij}]_{N\times N} \text{ and } \tilde{Y} = [\tilde{y}_{ij}]_{n \times n}$ as adjacency matrices describing the full and RDS network structures, respectively. \\
In this paper, we focus on un-directed networks only, such that
\begin{align*}
y_{ij} = y_{ji} = \begin{cases}
      1, & \text{if}\ \text{nodes } i, \text{and } j\text{ are connected in full network} \\
      0, & \text{otherwise,}
    \end{cases}
\end{align*}
\begin{align*}
\tilde{y}_{ij} = \tilde{y}_{ji} = \begin{cases}
      1, & \text{if}\ \text{nodes } i, \text{and } j\text{ are sampled and connected in the RDS sample} \\
      0, & \text{if node } i, \text{node } j\text{ are sampled, but not connected in the RDS sample.}
    \end{cases}
\end{align*}
The number of edges incident to a node is called the degree of that node. In Figure \ref{fig: figure 1}, each node has a degree at most 4. This is because RDS restricts each respondent's recruitment has to be no more than 3. This results in two types of degree for nodes in the RDS network sample, one is their degree in the RDS sample, and the other one is their degree in the hidden full network. For example, in the drug user RDS network, if person A is recruited as a sample, even though its degree in the RDS network is 3, its degree in the population might be greater than 3 because person A might know more than 3 drug users and he just recruited two or three of them into the sample. We denote the degree for node $i$ in the hidden full network as $d_i$. In this paper, when we use degree we mean degree in the population if not otherwise specified.\\
RDS data usually have node features describing each sample. We focus on clustering node-attributed RDS sample in this paper. Assume we have one continuous and one discrete feature describing the nodes. Without loss of generality, we label the sampled nodes with indices $1, \cdots, n$. Then,
\begin{itemize}
\item $X_1$ and $\tilde{X}_1$ are the continuous variables for the full and RDS networks, respectively. 
\item $X_2$ and $\tilde{X}_2$ are the discrete variables for the full and RDS networks, respectively. 
\item $Z = [z_{ik}]_{N\times K}$ and $\tilde{Z} = [\tilde{z}_{ik}]_{n \times K}$ are matrices describing latent cluster status for the attributed full and RDS networks. K is the number of latent clusters in the full network. 
\begin{align*}
z_{ik} = \begin{cases}
      1, & \text{if}\ \text{node i is in the } k^{th} \text{ cluster} \\
      0, & \text{otherwise,}
    \end{cases}
\end{align*}
\begin{align*}
\tilde{z}_{ik} = \begin{cases}
      1, & \text{if}\ \text{node i is in the } k^{th} \text{ cluster and is sampled} \\
      0, & \text{otherwise,}
    \end{cases}
\end{align*}
Note that our goal is to get latent group memberships for nodes in the RDS network sample, which reflect their group memberships in the full network, which is $z_{ik} = \tilde{z}_{ik}$ for node $i$ in the RDS network. Furthermore,
\item $S = [S_i]_{n\times 1}$ is the node sampling probability vector, where 
$$S_i = P(\textnormal{node i is sampled}).$$
\item $SS = [SS_{ij}]_{n \times n}$ is the node pair sampling probability matrix,
$$SS_{ij} = P(\textnormal{node i and node j are sampled}).$$
\item $R = [R_{ij}]_{n \times n}$ is the edge sampling probability matrix,
$$R_{ij} = P(\tilde{Y}_{ij}=1 | Y_{ij} = 1).$$
\end{itemize}
\subsection{Node and Node Pair Sampling Probabilities in RDS}
The sampling probability for each node is highly related with its degree in the population. Taking an extreme case as an example, when we sample drug users' networks using RDS, if drug user A knows zero other drug users, and drug user B is a drug dealer who knows many other drug users, then person B has much higher degree than drug user A and has much higher probability to be sampled than person A, because person B knows many more other drug users and is more likely to be recruited into the sample. Since we have node features describing each node in the RDS network, unequal node sampling probabilities also means that those node features are sampled unequally. Therefore, in order to get a log-likelihood representing the full network from node features of the sample, taking node sampling probabilities into consideration is necessary. \\
RDS is a without replacement sampling procedure, so node sampling probability is not simply proportional to its degree. Gile (2011) proposed successive sampling (SS) to get improved node sampling probabilities. By iterating the successive sampling procedure to approximate RDS, Gile (2011) mapped nodes with degree k to their sampling probabilities $S_k$ with $f: d\rightarrow S_k$. Following Gile (2011)'s node sampling probability, we can extend to get node pair sampling probabilities $SS_{kh}$ for node pairs with one node having degree k and the other having degree h, through $g: (k,h)\rightarrow SS_{kh}$. In the second step of estimating node sampling probabilities in Gile's (2011) paper, we can add estimating node pair sampling probabilities by 
$$g_{SS}((k,h);n, N^i) \approx \frac{U_k\cdot U_h+1}{M\cdot N_k^i\cdot N_h^i +1},$$
where $U_k, k = 1, \cdots, K$ is total number of observed units of size k in the M simulations. 
\subsection{Edge Sampling Probabilities in RDS}
In a RDS network sample, if two nodes are connected, they must also be connected in the population network. If they are not connected in the RDS network sample, they may still be connected in the population network because of the without replacement sampling property of RDS. Node connections or edges play an important role in network clustering, so reflecting a true connection underlying the RDS network is critical. Therefore, edge sampling is worth considering if we want to get population clustering of nodes from the RDS network.\\
Due to link-tracing and without replacement sampling, edge sampling probabilities are not uniform in RDS. Ott and Gile (2006) extended the successive sampling approximation to estimate edge sampling probabilities in RDS \citep{Ott2016}. Sampling probabilities are summarized below, 
\begin{align*}
    \text{Node pair sampling probability } SS_{ij} &= P(\text{i,j are sampled})\\
    & = P(\text{i,j are sampled} | Y_{ij} = 1)\\
    & = P(\text{i,j are sampled} | Y_{ij} = 0),
\end{align*}
\begin{align*}
      \text{Edge sampling probability } R_{ij} &= P(\text{i,j are sampled and connected in RDS}|Y_{ij}=1)\\
    &= P(\tilde{Y}_{ij}=1|Y_{ij}=1),
\end{align*}
\begin{align*}
    &P(\text{i,j are sampled and not connected in RDS}|Y_{ij}=1) \\
    &= P(\tilde{Y}_{ij}=0|Y_{ij}=1)\\
    & = P(\text{i,j are sampled}|Y_{ij}=1) - P(\text{i,j are sampled and connected in RDS}|Y_{ij}=1) \\
    & = P(\text{i,j are sampled}) - P(\tilde{Y}_{ij}=1|Y_{ij}=1)\\
    &= SS_{ij} - R_{ij},\\
    &P(\text{i,j are sampled and connected}|Y_{ij}=0)\\
    &= P(\tilde{Y}_{ij}=1|Y_{ij}=0)\\
    & = 0,\\
    &P(\text{i,j are sampled and not connected}|Y_{ij}=0)\\
    & = P(\tilde{Y}_{ij}=0|Y_{ij}=0)\\
    & = P(\text{i,j are sampled}|Y_{ij}=0) - P(\text{i,j are sampled and connected}|Y_{ij}=0)\\
    & = SS_{ij} - 0 \\
    & = SS_{ij},
\end{align*}
Overall, we can summarize edge sampling probabilities in the contingency table: 
\begin{table}[ht]
\begin{center}
\caption{Edge sampling probability}
    \label{tab: table 1}
    \begin{tabular}{|c|c|c|c|c|}
    \hline
        \diagbox[width=10em]{Full\\Network}{RDS\\Network} & $\tilde{Y}_{ij}=0$ &$\tilde{Y}_{ij}=1$ & (i,j) not sampled\\
         \hline
         $Y_{ij}=1$ & $(SS_{ij} - R_{ij})P(Y_{ij}=1)$ & $R_{ij}P(Y_{ij}=1)$ & $(1-SS_{ij})P(Y_{ij}=1)$\\
         \hline
           $Y_{ij}=0$ & $SS_{ij}P(Y_{ij}=0)$ & 0 & $(1-SS_{ij})P(Y_{ij}=0)$\\
          \hline
    \end{tabular}
\end{center}
\end{table}
\section{Mixture Model and Weighted log-likehood Mixture Model For Clustering Node Attributed RDS Network Data}\label{Model}
Mixture modeling is a widely used clustering method. Gaussian mixtures are used for clustering continuous variables. Stochastic block models are used for clustering social networks. In this paper, we build a mixture model on both node features and network structures by assuming conditional independence between them given the cluster membership.
\subsection{Mixture model}\label{Model_1.1}
Assuming conditional independence between the social network and node features given their community labels, we can build a mixture model for the full network:
\begin{align*}
(X_{i1}|z_i = k) &\sim N(\mu_k, \sigma_k),\\
(X_{i2}|z_i = k) &\sim \text{Cat}(\theta_{1k},\cdots, \theta_{Mk}),\\
(Y_{ij}|z_i =k, z_j = h) &\sim \text{Bernoulli}(\phi_{kh}),\\
z_i &\sim \text{Cat}(\lambda_1,\cdots,\lambda_K),
\end{align*}
where 
\begin{itemize}
\item $k,h = 1, \cdots, K$, K is the number of latent clusters in the population.
\item $\mu_k, \sigma_k$ are the mean and standard deviation of the continuous variable in the $k^{th}$ cluster. 
\item $\theta_{mk} = P(X_{i2}=m|z_i=k)$ is probability that discrete variable $X_{i2} = m$ given node $i$ in the $k^{th}$ cluster, for any $i=1, \cdots, N$, $M$ is the number of categories for discrete covariate $X_2$, 
$$\sum_{m=1}^M \theta_{mk} = 1.$$ 
\item $\phi_{kh}=P(Y_{ij}=1|z_i=k,z_j=h)$ is the probability that node $i$ and $j$ are connected given node $i$ in the $k^{th}$ cluster and node $j$ in the $h^{th}$ cluster.
\item $\lambda_k = P(z_i=k)$ is the probability that node $i$ is in the $k^{th}$ cluster, for any $i=1, \cdots, N$, 
$$\sum_{k=1}^K \lambda_k=1.$$
\end{itemize} 
If we ignore sampling, a naive approach is to apply the mixture model for the full network directly to the RDS network sample. We set it as our Benchmark Model: 
\begin{align*}
(\tilde{X}_{i1}|z_i = k) &\sim N(\mu_k, \sigma_k),\\
(\tilde{X}_{i2}|z_i = k) &\sim \text{Cat}(\theta_{1k},\cdots, \theta_{Mk}),\\
(\tilde{Y}_{ij}|z_i =k, z_j = h) &\sim \text{Bernoulli}(\phi_{kh}),\\
\tilde{z}_i &\sim \text{Cat}(\lambda_1,\cdots,\lambda_K).
\end{align*}
In this paper, we apply variational EM algorithm to do approximate maximum likelihood inference. This algorithm is applicable even for large networks with thousands of nodes \citep{Daudin2008}.\\
Given the above mixture model, the variational EM algorithm contains two steps, the variational E-step and the variational M-step. In the E-step of the traditional EM algorithm, we calculate the expectation of the full log-likelihood:
\begin{align*}
Q(\Theta|\Theta^{(t+1)}) &= E_{\tilde{Z}|\tilde{X}_1, \tilde{X}_2, \tilde{Y},\Theta^{(t)}}\text{log}L(\Theta;\tilde{X}_1, \tilde{X}_2, \tilde{Y},\tilde{Z}) \\
& = \sum_{i=1}^n \sum_{k=1}^K \pi_{ik}{\Large[}\text{log}P(\tilde{X}_{_{i1}}|z_{ik}) +\text{log}P(\tilde{X}_{_{i2}}|z_i=k)+\text{log}P(z_i=k){\Large]}\\
& +\frac{1}{2}\sum_{i,j=1,i\neq j}^n \sum_{k,h=1}^K \pi_{ik,jh}\text{log}P(\tilde{Y}_{ij}|z_i=k,z_j=h),
\end{align*}
where $\pi_{ik} = P(z_i=k|\tilde{X}_1, \tilde{X}_2, \tilde{Y}), \pi_{ik,jh} = P(z_i=k,z_j=h|\tilde{X}_1, \tilde{X}_2, \tilde{Y})$.\\
It is not easy to calculate $\pi_{ik}$ and $\pi_{ik,jh}$ because the cluster of node $i$ is not only associated with nodes connecting with it but is also dependent with other nodes not connecting with it. Considering this, the variational EM \citep{Daudin2008} is proposed by approximating $P(Z|\tilde{X}_1, \tilde{X}_2, \tilde{Y},\Theta^{(t)})$ with $R(Z) = \Pi_{i=1}^n \tau_{iz_i}$, where $\tau_{ik} \approx P(z_i=k|\tilde{X}_1, \tilde{X}_2, \tilde{Y},\Theta)$, $\tau_{ik,jh} = \tau_{ik}\tau_{jh} \approx P(z_i=k,z_j=h|\tilde{X}_1, \tilde{X}_2, \tilde{Y},\Theta)$, and $\sum_{k=1}^K\tau_{ik}=1$ for any $i = 1, \cdots, n$.
\begin{itemize}
\item The variational E-step: 
Modify the E-step of the traditional EM algorithm by approximating $\pi_{ik}$ with $\tau_{ik}$: 
\begin{align*}
\mathcal{Q}(\Theta|\Theta^{(t)}) &= E_{R(Z)}\text{log}L(\Theta;\tilde{X}_1, \tilde{X}_2, \tilde{Y}) - E_{R(Z)} D_{KL}(R(Z)||P(Z|\tilde{X}_1, \tilde{X}_2, \tilde{Y}))\\
& = E_{R(Z)}\text{log}L(\Theta;\tilde{X}_1, \tilde{X}_2, \tilde{Y},\tilde{Z}) - E_{R(Z)} \text{log}R(Z)\\
& = \sum_{i=1}^{n} \sum_{k=1}^{K} \tau_{ik}{\Large[}\text{log}P(\tilde{X}_{_{i1}}|z_{ik}) +\text{log}P(\tilde{X}_{_{i2}}|z_i=k)+\text{log}P(z_i=k){\Large]}\\
& +\frac{1}{2}\sum_{i,j=1,i\neq j}^{n}\sum_{k,h=1}^{K} \tau_{ik} \tau_{jh}\text{log}P(\tilde{Y}_{ij}|z_i=k,z_j=h) - \sum_{i=1}^{n}\sum_{k=1}^{K} \tau_{ik}\text{log}\tau_{ik},
\end{align*}
where $D_{KL}(R(Z)||P(Z|\tilde{X}_1, \tilde{X}_2, \tilde{Y})) = \sum_{Z} R(Z)\text{log}\frac{R(Z)}{P(Z|\tilde{X}_1, \tilde{X}_2, \tilde{Y})}$ is Kullback–Leibler (KL) divergence from $R(Z)$ to $P(Z|\tilde{X}_1, \tilde{X}_2, \tilde{Y})$, $D_{KL} \geq 0$. The closer it is to 0, the better $R(Z)$ approximates $P(Z|\tilde{X}_1, \tilde{X}_2, \tilde{Y})$.
\item The variational M-step: Similar to the M-step in the EM algorithm, in this step, we also update parameters by maximizing the expectation in the variational E-step. 
$$\Theta^{(t+1)} = \underset{\theta}{max}\mathcal{Q}(\Theta|\Theta^{(t)}),$$
Taking the derivative of $\mathcal{Q}(\Theta|\Theta^{(t)})$ for each parameter, in the $(t+1)^{th}$ iteration we update parameters with: 
\begin{align*}
\hat{\tau}_{ik}^{(t+1)} &\propto \hat{\lambda}_k^{(t)}P(\tilde{X}_{i1}|\hat{\mu}_k^{(t)},\hat{\sigma}_k^{(t)})P(\tilde{X}_{i2}|\hat{\theta}_{mk}^{(t)},m=1,...,M)\\
& \text{       }\Pi_{j\neq i}\Pi_{h=1}^K {\Large[}P(\tilde{Y}_{ij}|\hat{\phi}_{kh}^{(t)}){\Large]},\\
\hat{\lambda}_k^{(t+1)} &= \frac{\sum_{i=1}^n \hat{\tau}_{ik}^{(t+1)}}{n},\\
\hat{\mu}_{k}^{(t+1)} &= \frac{\sum_{i}\hat{\tau}_{ik}^{(t+1)}x_{i1}}{\sum_{i}\hat{\tau}_{ik}^{(t+1)}}, \text{  }\hat{\sigma^2}_{k}^{(t+1)} = \frac{\sum_{i}\hat{\tau}_{ik}^{(t+1)}(x_{i1} - \hat{\mu}_{k}^{(t+1)})^2}{\sum_{i}\hat{\tau}_{ik}^{(t+1)}},\\
\hat{\theta}_{mk}^{(t+1)} &= \frac{\sum_{i}\tau_{ik}^{(t+1)}\text{I}(X_{i2}==m)}{\sum_{i}\tau_{ik}^{(t+1)}},\\
\hat{\phi}_{kh}^{(t+1)} &= \frac{\sum_{i\neq j}\tau_{ik}^{(t+1)}\tau_{jh}^{(t+1)}\tilde{Y}_{ij}}{\sum_{i\neq j}\tau_{ik}^{(t+1)}\tau_{jh}^{(t+1)}}.
\end{align*}
\end{itemize}
\subsection{Weighted Log-likelihood Mixture model}\label{Model_1.2}
As we discussed in Section \ref{Background}, RDS results in non-uniform node and edge sampling probabilities and it's necessary to consider both of them for valid clustering results and parameters estimation. In the paper, we modify the log-likelihood in the mixture model in Section \ref{Model_1.1} by adding node and edge weights as the inverse of their sampling probabilities to approximate the log-likelihood in the underlying graph of the RDS network. Based on this weighted log-likelihood we can update parameters and find cluster membership for nodes in the underlying graph. We call this model the weighted log-likelihood mixture model.\\
Given the full network mixture model, for nodes $i,j=1,\cdots,N$:
\begin{align*}
(X_{i1}|z_i = k) &\sim N(\mu_k, \sigma_k),\\
(X_{i2}|z_i = k) &\sim \text{Cat}(\theta_{1k},\cdots, \theta_{Mk}),\\
(Y_{ij}=1|z_i =k, z_j = h) &\sim \text{Bernoulli}(\phi_{kh}),\\
Z_i &\sim \text{Cat}(\lambda_1,\cdots,\lambda_K),
\end{align*}
the variational E-step starts with: 
\begin{align*}
\mathcal{Q}_{full}(\Theta|\Theta^{(t)}) &= E_{R(Z)}\text{log}L(\Theta;X_1, X_2, Y) - E_{R(Z)} D_{KL}(R(Z)||P(Z|X_1, X_2, Y))\\
& = \sum_{i=1}^N \sum_{k=1}^K \tau_{ik}{\Large[}\text{log}P(X_{_{i1}}|z_{ik}) +\text{log}P(X_{_{i2}}|z_i=k)+\text{log}P(z_i=k){\Large]} && \cdot \dotsm \cdot\textbf{A}\\
& +\frac{1}{2}\sum_{i,j=1,i\neq j}^N \sum_{k,h=1}^K \tau_{ik} \tau_{jh}{\Large[}Y_{ij}\text{log}P(Y_{ij}=1|z_i=k,z_j=h)\Large] && \cdot \dotsm \cdot\textbf{B}\\
& + \frac{1}{2}\sum_{i,j=1,i\neq j}^N \sum_{k,h=1}^K \tau_{ik} \tau_{jh}{\Large[}(1 -Y_{ij})\text{log}P(Y_{ij}=0|z_i=k,z_j=h)\Large] && \cdot \dotsm \cdot\textbf{C}\\
&- \sum_{i=1}^N\sum_{k=1}^K \tau_{ik}\text{log}\tau_{ik}. && \cdot \dotsm \cdot\textbf{D}
\end{align*}
In $\mathcal{Q}_{full}(\Theta|\Theta^{(t)})$, the full network log-likelihood contains four parts, part A is the log-likelihood of node features, part B is the log-likelihood of two connected nodes, part C is the log-likelihood of two nodes not connected, and part D is the penalty term from the KL divergence. \\
Based on node sampling probabilities $S = \{S_i, i=1,\cdots,n\}$, part A can be approximated by weighted log-likelihood from node features in the RDS network: 
$$\text{part A} \approx \sum_{i=1}^n \sum_{k=1}^K \tau_{ik}\frac{1}{S_i}{\Large[}\text{log}P(X_{_{i1}}|z_{ik}) +\text{log}P(X_{_{i2}}|z_i=k)+\text{log}P(z_i=k){\Large]},$$
Part D can be approximated using node sampling probabilities as well: 
$$\text{part D} \approx \sum_{i=1}^n \sum_{k=1}^K \tau_{ik}\frac{1}{S_i}\tau_{ik}\text{log}\tau_{ik},$$
Since all edges in the RDS network are sampled from edges in the full network with sampling probabilities $R = {R_{ij, i,j=1, \cdots,n}}$ and $R_{ij} = P(\tilde{Y}_{ij}=1|Y_{ij}=1)$, part B can be approximated by weighted log-likelihood of edges in the RDS network: 
$$\text{part B} \approx \sum_{i,j=1,i\neq j}^n \sum_{k,h=1}^K \tau_{ik} \tau_{jh}\frac{1}{R_{ij}}{\Large[}\tilde{Y}_{ij}\text{log}P(Y_{ij}=1|z_i=k,z_j=h)\Large].$$
Two nodes not connected in the RDS network may still be connected in the full network. To approximate part C, we first need to estimate the probability that un-connected nodes in the sample are also not connected in the full network, denoted by $P(Y_{ij}=0|\tilde{Y}_{ij}=0)$: 
\begin{align*}
&P(Y_{ij}=0|\tilde{Y}_{ij} = 0)\\
& = \frac{P(Y_{ij}=0, \tilde{Y}_{ij}=0)}{P(\tilde{Y}_{ij}=0)}\\
&= \frac{P(Y_{ij}=0, \tilde{Y}_{ij}=0)}{P(Y_{ij}=0, \tilde{Y}_{ij}=0) + P(Y_{ij}=1, \tilde{Y}_{ij}=0)}\\
& = \frac{P(\tilde{Y}_{ij}=0|Y_{ij}=0)P(Y_{ij}=0)}{P(\tilde{Y}_{ij}=0|Y_{ij}=0)P(Y_{ij}=0) + P(\tilde{Y}_{ij}=0|Y_{ij}=1)P(Y_{ij}=1)}\\
     &=   \frac{SS_{ij}P(Y_{ij}=0)}{SS_{ij}P(Y_{ij}=0) + (SS_{ij}-R_{ij})P(Y_{ij}=1)} \\
    &= \frac{SS_{ij}P(Y_{ij}=0)}{SS_{ij} - R_{ij}P(Y_{ij} = 1)}.
\end{align*}
Assume sampling probabilities are independent given cluster labels. We have $P(Y_{ij}=0|\tilde{Y}_{ij} = 0,z_i=k,z_j=h) = \frac{SS_{ij}P(Y_{ij}=0|z_i=k,z_j=h)}{SS_{ij} - R_{ij}P(Y_{ij} = 1|z_i=k,z_j=h)}$.
Meanwhile, from Table \ref{tab: table 1} we also have sampling probabilities of two unconnected nodes, $P(\tilde{Y}_{ij}=0|Y_{ij}=0) = SS_{ij}$. Then we can approximate part C by:
\begin{align*}
    &\text{part C}  \\
    \approx &\sum_{i,j=1,i\neq j}^n \sum_{k,h=1}^K \tau_{ik} \tau_{jh}\frac{1}{SS_{ij}}{\Large[}P(Y_{ij}=0|\tilde{Y}_{ij}=0,z_i=k,z_j=h)(1 -\tilde{Y}_{ij})\text{log}P(Y_{ij}=0|z_i=k,z_j=h){\Large]} \\
    = &\sum_{i,j=1,i\neq j}^n \sum_{k,h=1}^K \tau_{ik} \tau_{jh}\frac{1}{SS_{ij}}{\Large[}\frac{SS_{ij}P(Y_{ij}=0|z_i=k,z_j=h)}{SS_{ij} - R_{ij}P(Y_{ij} = 1|z_i=k,z_j=h)}(1 -\tilde{Y}_{ij})\text{log}P(Y_{ij}=0|z_i=k,z_j=h){\Large]} \\
    = &\sum_{i,j=1,i\neq j}^n \sum_{k,h=1}^K \tau_{ik} \tau_{jh}{\Large[}(1 -\tilde{Y}_{ij})\frac{P(Y_{ij}=0|z_i=k,z_j=h)\text{log}P(Y_{ij}=0|z_i=k,z_j=h)}{SS_{ij} - R_{ij}P(Y_{ij} = 1|z_i=k,z_j=h)}{\Large]}.
\end{align*}
With all these weights, we get the full log-likelihood approximation for the variational E-step: 

\begin{align*}
&\mathcal{Q}_{full}(\Theta|\Theta^{(t)}) = \text{part A + part B + part C - part D}\\
& \approx \mathcal{Q}_{w}(\Theta|\Theta^{(t)})\\
& = \sum_{i=1}^n \sum_{k=1}^K \tau_{ik}\frac{1}{S_i}{\Large[}\text{log}P(X_{_{i1}}|z_{ik}) +\text{log}P(X_{_{i2}}|z_i=k)+\text{log}P(z_i=k){\Large]} && \cdot \dotsm \cdot\textbf{w-A}\\
& +\frac{1}{2}\sum_{i,j=1,i\neq j}^n \sum_{k,h=1}^K \tau_{ik} \tau_{jh}\frac{1}{R_{ij}}{\Large[}\tilde{Y}_{ij}\text{log}P(Y_{ij}=1|z_i=k,z_j=h){\Large]} && \cdot \dotsm \cdot\textbf{w-B} \\
& + \frac{1}{2}\sum_{i,j=1,i\neq j}^n \sum_{k,h=1}^K \tau_{ik} \tau_{jh}{\Large[}(1 -\tilde{Y}_{ij})\frac{P(Y_{ij}=0|z_i=k,z_j=h)\text{log}P(Y_{ij}=0|z_i=k,z_j=h)}{SS_{ij} - R_{ij}P(Y_{ij} = 1|z_i=k,z_j=h)}{\Large]} && \cdot \dotsm \cdot\textbf{w-C} \\
&- \sum_{i=1}^n\sum_{k=1}^K \frac{1}{S_i}\tau_{ik}\text{log}\tau_{ik} && \cdot \dotsm \cdot\textbf{w-D}\\
& = \sum_{i=1}^n\sum_{k=1}^K \tau_{ik}\frac{1}{S_i}{\Large[}\text{log}(\frac{1}{2\sigma_k\sqrt{2\pi}}) - \frac{(x_{i1} - \mu_k)^2}{2\sigma_k^2} + \text{log}\sum_{m=1}^M I\{x_{i2}==m\}\theta_{mk} + \text{log}\lambda_k{\Large]}\\
& + \frac{1}{2}\underset{i,j=1,\cdots,n;i\neq j}{\sum}\sum_{k,h=1}^K \tau_{ik}\tau_{jh}{\Large[}\tilde{Y}_{ij}\frac{\text{log}\phi_{kh}}{R_{ij}}
+ (1 - \tilde{Y}_{ij})(1-\phi_{kh})\frac{\text{log}(1-\phi_{kh})}{SS_{ij} - R_{ij}\phi_{kh}}{\Large]}\\
& - \sum_{i=1}^n\sum_{k=1}^K \frac{1}{S_i}\tau_{ik}\text{log}\tau_{ik}.
\end{align*}
In the variational M-step, we update parameters by maximizing the weighted log-likelihood in the variational E-step:
$$\Theta_w^{(t+1)} = \underset{\theta}{max}\mathcal{Q}_w(\Theta|\Theta^{(t)}),$$

\begin{align*}
\hat{\tau}_{ik}^{(t+1)} &\propto {\Large[}\hat{\lambda}_k^{(t)}P(X_{i1}|\hat{\mu}_k^{(t)},\hat{\sigma}_k^{(t)})P(X_{i2}|\hat{\theta}_{mk}^{(t)},m=1,...,M){\Large]}\Pi_{j\neq i}\Pi_{h=1}^K {\Large[}P(\tilde{Y}_{ij}|\hat{\phi}_{kh}^{(t)}){\Large]}^{\tau_{jh}^{(t)}S_i}\\ 
& = {\Large[}\hat{\lambda}_k^{(t)}P(X_{i1}|\hat{\mu}_k^{(t)},\hat{\sigma}_k^{(t)})P(X_{i2}|\hat{\theta}_{mk}^{(t)},m=1,...,M){\Large]}\\
& \text{     }\Pi_{j\neq i}\Pi_{h=1}^K {\Large[}(\hat{\phi}_{kh}^{(t)})^{\tilde{Y}_{ij}/R_{ij}}(1-\hat{\phi}_{kh}^{(t)})^{(1-\tilde{Y}_{ij})(1-\hat{\phi}_{kh}^{(t)})/(SS_{ij} - R_{ij}\hat{\phi}_{kh}^{(t)})}{\Large]}^{\tau_{jh}^{(t)}S_i},\\
\hat{\lambda}_k^{(t+1)} &= \frac{\sum_{i=1}^n \hat{\tau}_{ik}^{(t+1)}/S_i}{n},\\
\hat{\mu}_{k}^{(t+1)} &= \frac{\sum_{i}\hat{\tau}_{ik}^{(t+1)}/S_ix_{i1}}{\sum_{i}\hat{\tau}_{ik}^{(t+1)}/S_i}, \text{  }\hat{\sigma^2}_{k}^{(t+1)} = \frac{\sum_{i}\hat{\tau}_{ik}^{(t+1)}/S_i(x_{i1} - \hat{\mu}_{k}^{(t+1)})^2}{\sum_{i}\hat{\tau}_{ik}^{(t+1)}/S_i},\\
\hat{\theta}_{mk}^{(t+1)} &= \frac{\sum_{i}\tau_{ik}^{(t+1)}/S_i\text{I}(X_{i2}==m)}{\sum_{i}\tau_{ik}^{(t+1)}/S_i},\\
\frac{\partial \mathcal{Q}_w}{\partial \phi_{k,h}^{(t+1)}} = &\sum_{i,j=1\cdots,n;i\neq j} \tau_{ik}^{(t+1)}\tau_{jh}^{(t+1)} [\frac{\tilde{Y}_{ij}}{R_{ij}\phi_{k,h}^{(t+1)}} + \\
& (1 - \tilde{Y}_{ij})\frac{(R_{ij}-S_{ij})\text{log}(1 - \phi_{k,h}^{(t+1)}) - (S_{ij} - R_{ij}\phi_{k,h}^{(t+1)})}{(S_{ij}-R_{ij}\phi_{k,h}^{(t+1)})^2}].\\
\text{Set }\frac{\partial \mathcal{Q}_w} \partial\phi_{k,h}^{(t+1)} & = 0, \text{ and we can solve for } \phi_{k,h}^{(t+1)} \text{ using Newton-Raphson iteration.}
\end{align*}

\subsection{Weighted log-likelihood mixture model with tuning parameter}\label{Modle_1.3}
In the weighted log-likelihood mixture model, the full log-likelihood approximation is
\begin{align*}
\mathcal{Q}_w(\Theta|\Theta^{(t)}) = \text{part } \textbf{w-A} + \text{part } \textbf{w-B} + \text{part } \textbf{w-C} - \text{part } \textbf{w-D},
\end{align*}
where part \textbf{w-A} is the weighted log-likelihood from covariates, and (part \textbf{w-B} + part \textbf{w-C}) is the weighted log-likelihood from the network structure. In this section, we add a tuning parameter to balance contribution of the network structure and covariates, where 
\begin{align*}
\mathcal{Q}_{w;\alpha}(\Theta|\Theta^{(t)}) = \text{part } \textbf{w-A} + \alpha * (\text{part } \textbf{w-B} + \text{part } \textbf{w-C}) - \text{part } \textbf{w-D}.
\end{align*}
When $\alpha=0$, the clustering is based on covariates only, when $\alpha=1$, $\mathcal{Q}_{w;\alpha}(\Theta|\Theta^{(t)}) = \mathcal{Q}_{w}(\Theta|\Theta^{(t)})$, larger $\alpha$, contribution of the network structure is larger. This is similar to spectral clustering with covariates (\citep{Binkiewicz2017}\citep{Shiga2007}). 
Adding the tuning parameter $\alpha$ only effects the cluster memberships of nodes.
\begin{align*}
\hat{\tau}_{ik;\alpha}^{(t+1)} &\propto {\Large[}\hat{\lambda}_k^{(t)}P(X_{i1}|\hat{\mu}_k^{(t)},\hat{\sigma}_k^{(t)})P(X_{i2}|\hat{\theta}_{mk}^{(t)},m=1,...,M){\Large]}\Pi_{j\neq i}\Pi_{h=1}^K {\Large[}P(\tilde{Y}_{ij}|\hat{\phi}_{kh}^{(t)}){\Large]}^{\alpha\tau_{jh}^{(t)}S_i}\\
    & = {\Large[}\hat{\lambda}_k^{(t)}P(X_{i1}|\hat{\mu}_k^{(t)},\hat{\sigma}_k^{(t)})P(X_{i2}|\hat{\theta}_{mk}^{(t)},m=1,...,M){\Large]}\\
& \text{     }\Pi_{j\neq i}\Pi_{h=1}^K {\Large[}(\hat{\phi}_{kh}^{(t)})^{\tilde{Y}_{ij}/R_{ij}}(1-\hat{\phi}_{kh}^{(t)})^{(1-\tilde{Y}_{ij})(1-\hat{\phi}_{kh}^{(t)})/(SS_{ij} - R_{ij}\hat{\phi}_{kh}^{(t)})}{\Large]}^{\alpha\tau_{jh}^{(t)}S_i}.
\end{align*}
Updates for all the other parameters are the same as those of the mixture model with weighted log-likelihood in Section \ref{Model_1.2}.
\section{Clustering evaluation and tuning parameter selection}\label{Evaluation}
When both node features and network have communities, we need to decide the tuning parameter value $\alpha$ to get desired clusters. To check if the clustering is what we want for the network with node attributes, we need to evaluate the clustering quality in terms of network structure and in terms of node attributes. Then the tuning parameter $\alpha$ can be chosen based on clustering evaluation metrics. \\
Evaluating the quality of clustering algorithms is typically in two ways, internal evaluation and external evaluation. The internal evaluation uses a score to summarize clustering quality and the external evaluation compares a known classification in the data with the clustering got from the model. Popular internal evaluation metrics for network clustering include modularity, conductance, coverage \citep{Newman2006}\citep{Elmqvist2014}\citep{Schaeffer2007} and common internal evaluations for attributes are Silhouette index, Dunn’s indices, Davies-Bouldin index, etc \citep{Rousseeuw1987} \citep{Dunnt1974}\citep{Davies1979}. Popular external clustering evalution metrics include purity, entropy, normalized mutual information, F measure, Rand index \citep{Larsen1999}\citep{Strehl2003}\citep{Rendon2011}. In this paper, we focus on modularity for the network clustering evaluation and normalized mutual information for evaluating clustering of node features. For both of them, larger value indicates better clustering, can be used to compare different clustering algorithms and choose number of clusters for the clustering algorithm. In this paper, we use these two clustering evaluation metrics to determine tuning parameter $\alpha$ as well.\\
Modularity evaluates the strength of division of a network into clusters. Assume network G is clustered into K clusters with vertex sets $C = \{C_1, \cdots, C_K\}$, then the modularity $\textnormal{Q}(C)$ is
\begin{align*}
    \textnormal{Q}(C) = \sum_{k=1}^K e_{kk} - a_k^2, 
\end{align*}
where $E_{kl} = \sum_{i \neq j} (\tilde{Y}_{ij}|z_i=K,z_j=l)$, $e_{kk} = \frac{E_{kk}}{\sum_{k,l}E_{kl}}$ is fraction of edges with both vertices in cluster $k$. $a_k = \frac{\sum_{l}E_{kl}}{\sum_{k,l}E_{kl}}$ is the fraction of ends of edges incident to cluster k, $a_k^2$ is the expected fraction of edges with both vertices in cluster k if edges were randomly distributed. The range of modularity is [-1, 1]. Higher modularity means more edges are within clusters than between clusters.\\
Mutual Information measures mutual dependence between two random variables, $X$ and $C$: 
$$I(X,C) = \sum_{x}\sum_{c} p(x,c)\textnormal{log}\frac{p(x,c)}{p(x)p(c)}.$$
The Normalized Mutual Information (NMI) is: 
$$\textnormal{NMI}(X,C) = \frac{I(X,C)}{\sqrt{H(X)H(C)}},$$
where $\textnormal{NMI}(X,C) \in [0,1]$, $\textnormal{NMI}(X,C) = 0$ indicates $X$ and $C$ are independent, and larger NMI means better clustering. $H(X) = -\sum_{x}p(x)\textnormal{log}p(x)$ is entropy of $X$. It is also true that $I(X,C)= H(X) + H(C) - H(X,C)= H(X) - H(X|C) = H(C) - H(C|X)$.\\
In our dataset, we have continuous and discrete node features. To calculate the NMI for all features, we have three steps. Step 1, we cut the continuous variables into discrete variables. Step 2, we calculate NMI for each node feature. Step 3, we take average of NMIs got in step 2 as our final NMI for node features.\\
Since RDS gives an incomplete social network, we don't know $e_{kk}$ and $a_k$ for the full network. Fortunately, we can estimate them through sampling weights, 
$$\hat{e}_{kk} = \frac{\hat{E}_{kk}}{\sum_{k,l} \hat{E}_{kl}},$$
$$\hat{a}_k = \frac{\hat{E}_{kk} + \sum_{l\neq k} \hat{E}_{kl}}{\sum_{k,l} \hat{E}_{kl}},$$
where $\hat{E}_{kl} = \sum_{i\neq j}\frac{\tilde{Y}_{ij}}{R_{ij}}I(z_i=k,z_j=l)$, then $\hat{Q}(C) = \sum_{k} \hat{e}_{kk} - (\hat{a}_k)^2$. \\
We can also estimate NMI$(X,C)$ for the full network $\hat{\textnormal{NMI}}(X,C) = \frac{\hat{I}(X,C)}{\sqrt{\hat{H}(X)\hat{H}(C)}}$ with 
$$\hat{H}(X) = -\sum_{x}\hat{p}(x)\textnormal{log}\hat{p}(x),\text{ } \hat{p}(x) = \frac{\sum_{i}I(X_i = x)/S_i}{\sum_{i}1/S_i},$$
similarly, we can estimate $\hat{H}(C)$ and $\hat{H}(X|C)$. \\
By looking at how the clustering evaluation metrics, normalized mutual information $\hat{NMI}$ and modularity $\hat{Q}$ change with different values of $\alpha$, we can decide the best tuning parameter $\alpha$. 
\section{Simulation Study}\label{Simulation}
In this section, we compare clustering performance using the mixture model with and without weighted log-likelihood and with different values of tuning parameters in four different cases. For each case, we simulate 100 full networks with one continuous variable and one categorical variable, then we sample a RDS network from each full network. Finally, we apply the candidate mixture models on those RDS networks. A summary of the different cases is in Table \ref{tab: table 2}.

\begin{table*}
    \centering
    \caption{Parameters for different simulation cases. $\phi$ is parameter for the network connection, $\mu$ is mean of the continuous variable, $\theta$ is parameter for the categorical variable, $\lambda$ is parameter for the cluster membership.}
    \label{tab: table 2}
    \smallskip\noindent
    \scalebox{0.8}{
        \begin{tabular}{|p{5cm}|c| c |c |c|}
        \hline
         & $\phi$ & $\mu$ & $\theta$ & $\lambda$ \\
         \hline
        Case I: Both separate well & \begin{math}
        \phi = \begin{bmatrix} 
        0.1& 0.02\\
        0.02 & 0.2
        \end{bmatrix}
        \end{math} & [-2,2] & \begin{math}
        \theta = \begin{bmatrix} 
        0.8& 0.4\\
        0.2 & 0.6
        \end{bmatrix}
        \end{math} & 1/3 \\
        \hline  
        Case II: Features separate well, Network does not  & $ \phi = \begin{bmatrix} 
        0.05& 0.05\\
        0.05 & 0.05
        \end{bmatrix}$ & [-2,2] & $ \theta = \begin{bmatrix} 
        0.8& 0.4\\
        0.2 & 0.6
        \end{bmatrix}$  &1/3\\
        \hline  
        Case III: Network separates well, Features do not & $ \phi = \begin{bmatrix} 
        0.1& 0.02\\
        0.02 & 0.2
        \end{bmatrix}$ & [0,0] & $ \theta = \begin{bmatrix} 
        0.5& 0.5\\
        0.5 & 0.5
        \end{bmatrix}$  &1/3\\
        \hline  
        Case IV: Both do not separate well & $ \phi = \begin{bmatrix} 
        0.05& 0.05\\
        0.05 & 0.05
        \end{bmatrix}$  & [0,0] & $\theta = \begin{bmatrix} 
        0.5& 0.5\\
        0.5 & 0.5
        \end{bmatrix}$  & 1/3\\
        \hline  
        \end{tabular}
    }
\end{table*}

The full networks are generated by:
\begin{align*}
    G &\sim \text{SBM}(N = 600, \phi = \phi, \text{block.size}=c(200,400)),\\
    (X_{i1}|z_i=k) &\sim N(\mu_k,1), \text{ }k=1,2; i = 1,\cdots,600,\\
    (X_{i2}|z_i=k) &\sim \text{Cat}(\theta_{1k},\theta_{2k}), \text{ } k=1,2; i = 1,\cdots,600,
\end{align*}
where $\text{SBM}(N = 600, \phi = \phi, \text{block.size}=c(200,400))$ is a stochastic block model with size $N=600$, two blocks or communities of size 200 and 400. The social connection parameter within and between blocks is denoted by $\phi$. \\
The RDS network sample is obtained by RDS sampling from the complete network G with 5 seeds for $n=300$, 3 seeds for $n=100$ and 3 coupons for each node. The distribution of number of recruitments for each sample is $[0, 1, 2, 3]$ with probabilities of $[0.1, 0.2, 0.3, 0.4]$ respectively. 
\begin{figure*}
    \centering
    \caption{Full network and one RDS network sampled from it}
    \label{fig: figure 1}
    \includegraphics[scale=0.5]{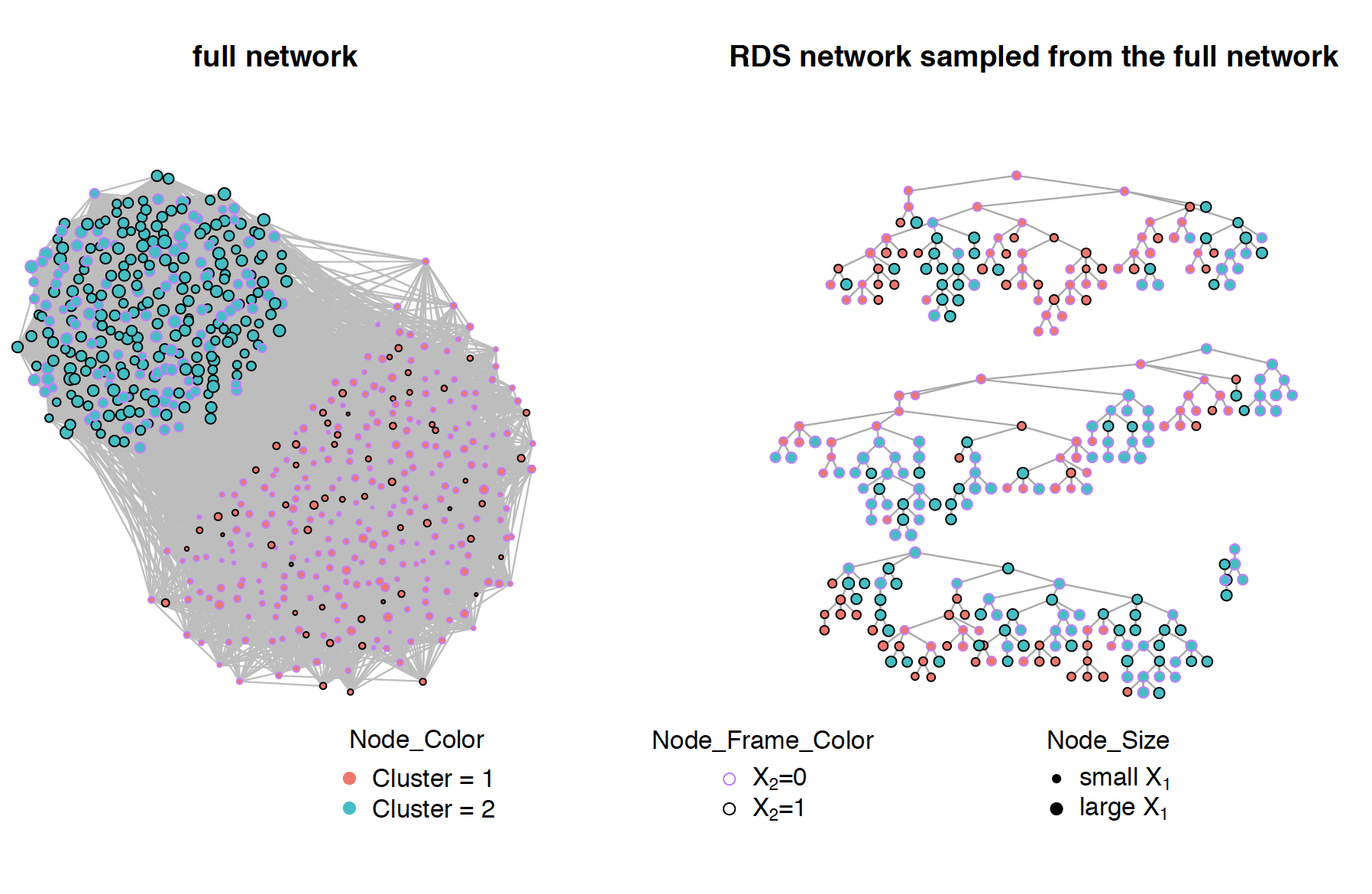}
\end{figure*}
One example of the full network and its sampled RDS network is plotted in Figure \ref{fig: figure 1}. In both networks, nodes are colored by their cluster labels, frame colored by their categorical values and sized by their continuous variable values. In this full network, both features and network structure separate well. We can see from the full and RDS network that people in the same cluster have similar node features and are more likely to connect. In the RDS network sample, nodes in different trees may be in the same cluster even though they come from different seeds in the RDS network sample and are not connected visually. To detect this latent clustering truth, node features play an important role. From the RDS network sample, we can also see that sampled degree for all nodes is at most 4 which is the maximum number of coupons each person can distribute plus 1.\\
In the simulation study, we take a full network of size $N=600$ and consider two types of its RDS sample with node samples of $n=300$ and $n=100$. Figures \ref{fig:figure 3} are plots of modularity and NMI with different values of tuning parameter $\alpha$, based on which we can determine the best tuning parameter for the corresponding RDS sample. Figures \ref{fig:figure 7} to \ref{fig:figure 10} are boxplots for parameter estimation, number of mis-clusterings, modularity and normalized mutual information by using five different models for the four different cases when $n=300$ and Table \ref{tab:table_MSE} summarizes parameter estimates under different models for RDS sample data with $n=300$ and $n=100$. Five different models we use are mixture model without weighting and $\alpha = 0$ (noW-alpha=0), mixture model without weighting and $\alpha=1$ (noW-alpha=1), mixture model with weighting and $\alpha = 0$ (W-alpha=0), mixture model with weighting and the best tuning parameter (W-alpha-star, alpha-star or $\alpha^*$ is the best selected tuning parameter for each RDS sample, e.g. we can see from plots in the first row of Figure \ref{fig:figure 3}, $\alpha^*=0.025$ for the first RDS sample (i=1) in case I.) and mixture model with weighting and $\alpha=1$ (W-alpha=1).
\subsection{Tuning parameter selection}
\begin{figure*}
    \centering
    \caption{Plots of Modularity and NMI vs Tuning parameter $\alpha$ in mixture model with weights for case I (both separate well); $i$ is the sample number, e.g. $i=1$ represents the first sampled RDS data.}
    \label{fig:figure 3}
    \includegraphics[scale=0.3]{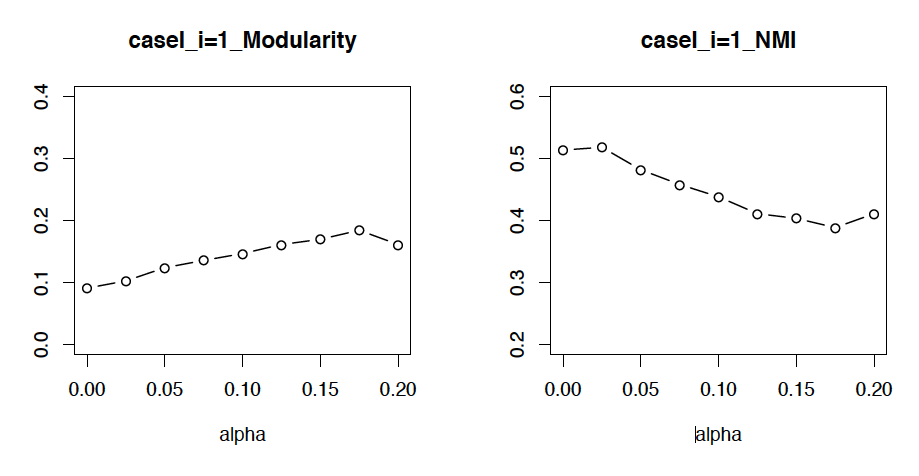}
    \includegraphics[scale=0.3]{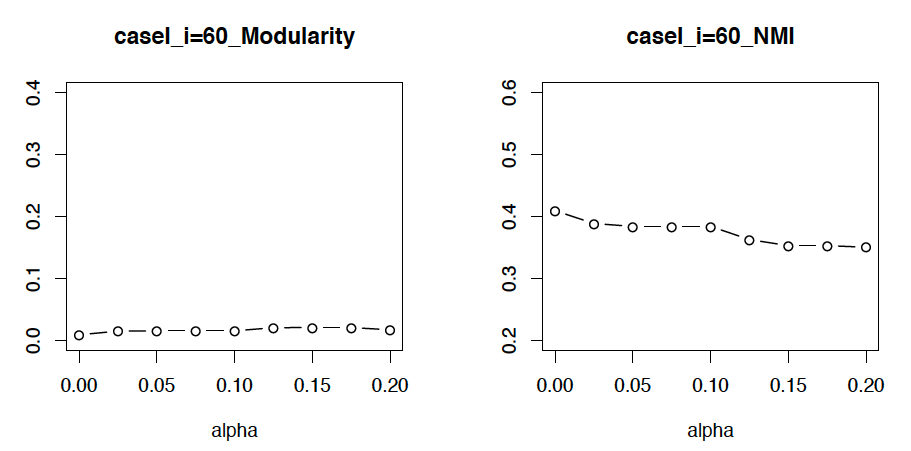}
    \includegraphics[scale=0.3]{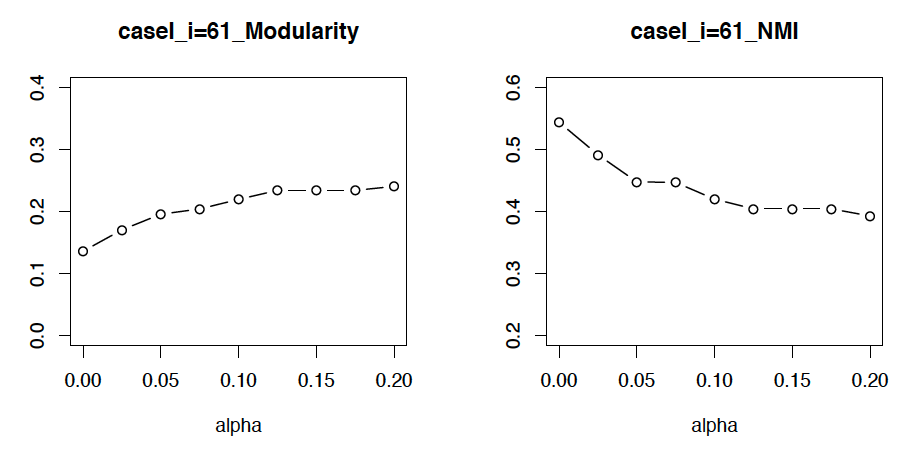}
    \includegraphics[scale=0.3]{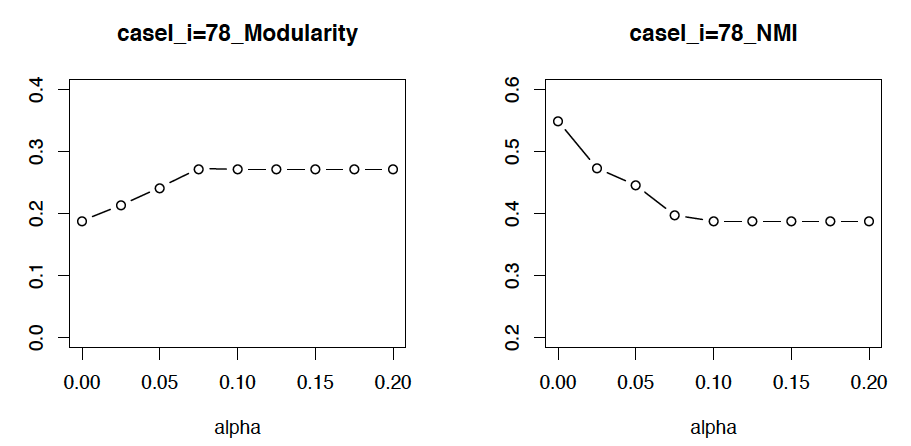}
\end{figure*}
The tuning parameter $\alpha$ controls contribution of network structure to the node cluster membership as we discussed in Section \ref{Modle_1.3}. In Figure \ref{fig:figure 3}, we plot modularity and NMI vs $\alpha$ for the first RDS sample ($i=1$) data, the $60^{th}$ RDS sample data, the $61^{th}$ and the $78^{th}$ RDS sample data sampled from the full network in case I. Plots for those four different sampled data differs obviously. It tells us that we need to find the best $\alpha$ for each RDS sample data even though they are sampled from the same full network. It's not hard to understand the reason why those plots are different, because RDS sample data may differ much even though they are from the same full population. We'll go through plots for those four RDS sampled data to discuss different situations in tuning parameter selection.\\
In the fist RDS sample data ($i=1$), we can see that both modularity and NMI are not small. This means that there are communities in the network structure and node features. When $\alpha=0$, the modularity is not small even though the clustering is based on node features only as we discussed in \ref{Modle_1.3}. This indicates that communities in the network and node features have overlaps. Moreover, the modularity has an increasing trend with increasing $\alpha$ and the NMI has a decreasing trend. The NMI has a larger decreasing speed after $\alpha=0.025$ and the modularity increases more from $\alpha=0$ to $\alpha=0.025$. Therefore, we choose $\alpha=0.025$ as our preferred tuning parameter value for the first sampled RDS data. \\
For the $i=60^{th}$ RDS sample data, we can see that the modularity is close to 0. This tells us that the network structure does not separate well in this sampled data. We pick $\alpha=0$ as the best tuning parameter so that we find the best communities in node features. \\
For the $i=61^{th}$ RDS sample data, the modularity has an increasing trend and NMI has a decreasing trend. It's hard to decide the best tuning parameter for this sample. If we choose the best $\alpha$ using the same procedure as we did for the first sampled data, we'll choose $\alpha=0$. However, it is also reasonable to pick $\alpha=0.025$ and $\alpha=0.05$ if we want to sacrifice some node feature information and get better community result for the network structure. One way to assist our selection of the tuning parameter, we can plot NMI for each node feature vs $\alpha$. Then we can select the best tuning parameter based on node features we care more about or node features that contribute more to the community detection.\\
For the $i=78^{th}$ RDS sample data, it's similar as the $61^{th}$ RDS sample data. But the modularity and NMI become almost flat after $\alpha=0.075$. This tells us that the clustering result won't change much anymore even though $\alpha$ increases or the contribution of network structure increases. This situation indicates that the clustering result is based on network structure only after $\alpha=0.075$ for this sampled data or the contribution of the network structure overtakes node features so much that the contribution of node features is negligible.\\
Similarly, we can choose the best tuning parameter $\alpha^*$ for each RDS sample data for each case. 
\subsection{Clustering evaluation and parameter estimation}
\begin{figure*}
    \centering
    \caption{Parameter estimations, Number of mis-clusterings, Modularity and NMI by using different models for case I (both separate well) when n=300}
    \label{fig:figure 7}
    \includegraphics[scale=0.4]{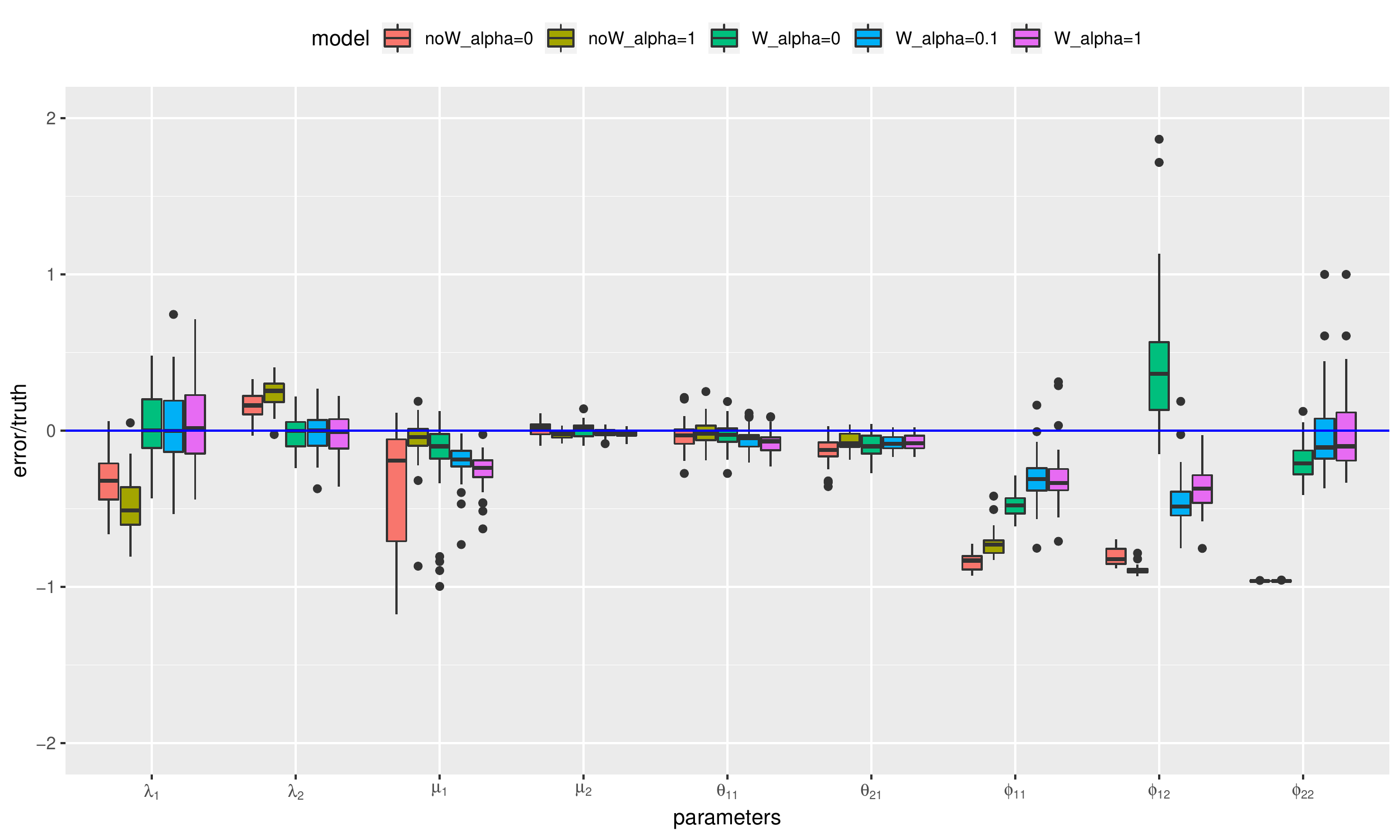}
    \includegraphics[scale=0.4]{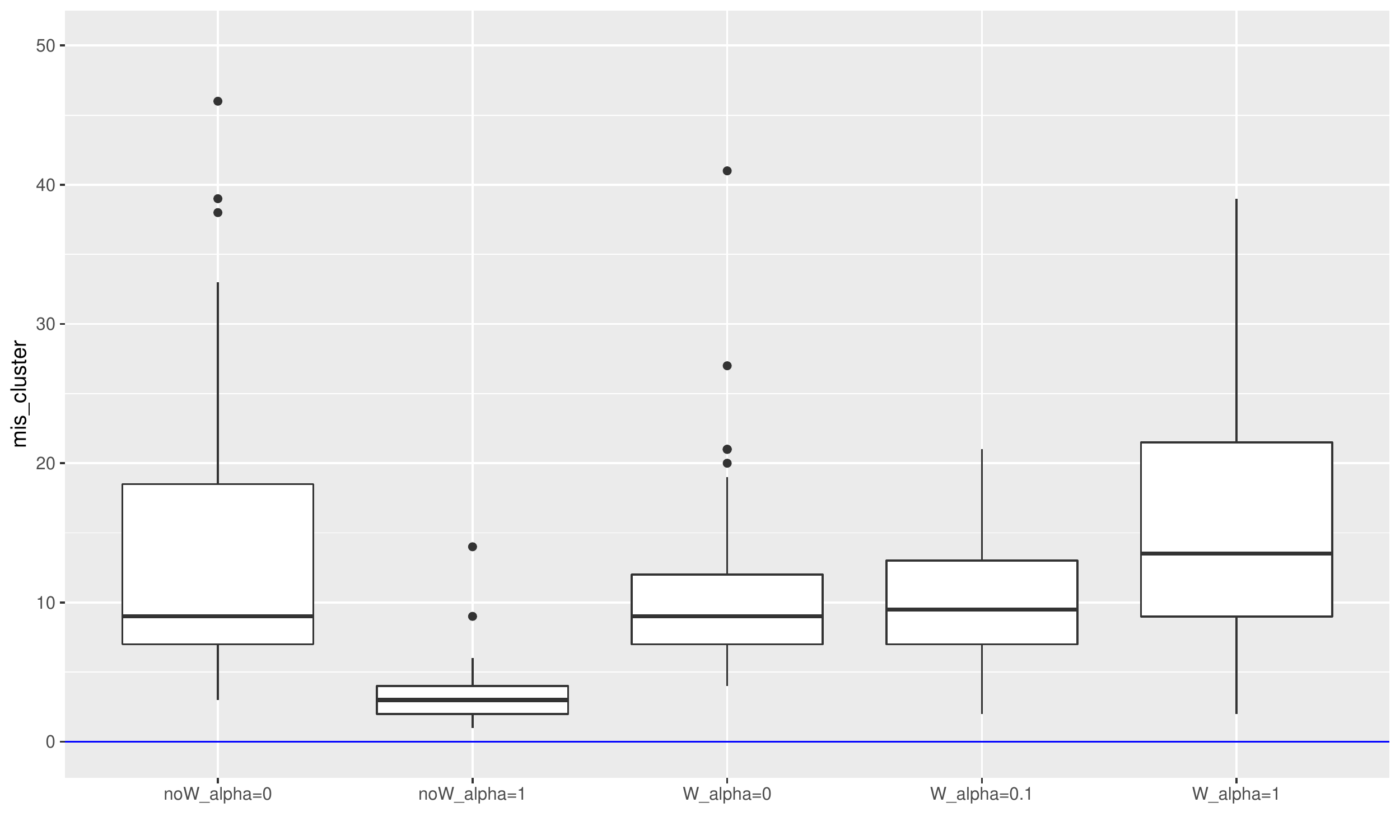}
    \includegraphics[scale=0.24]{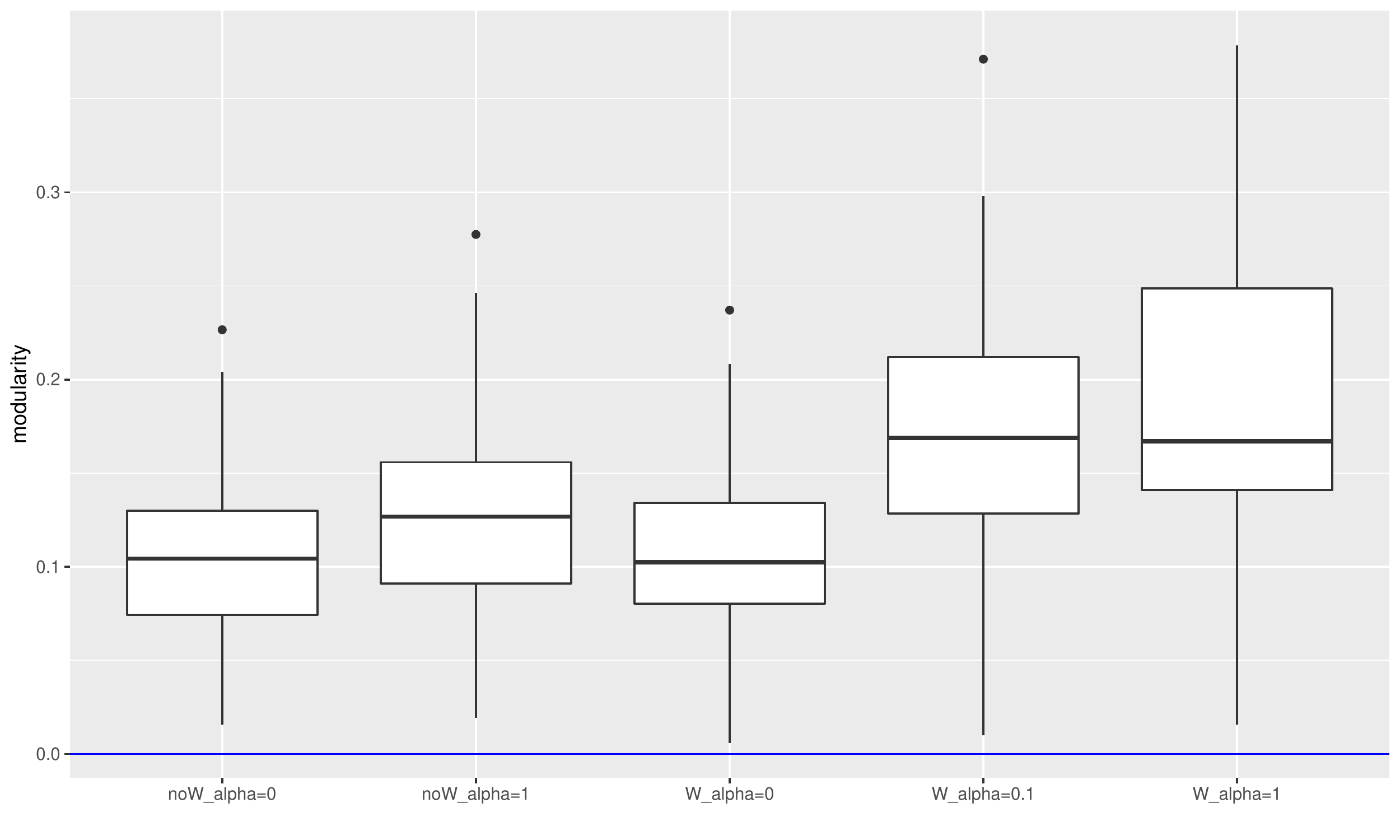}
    \includegraphics[scale=0.24]{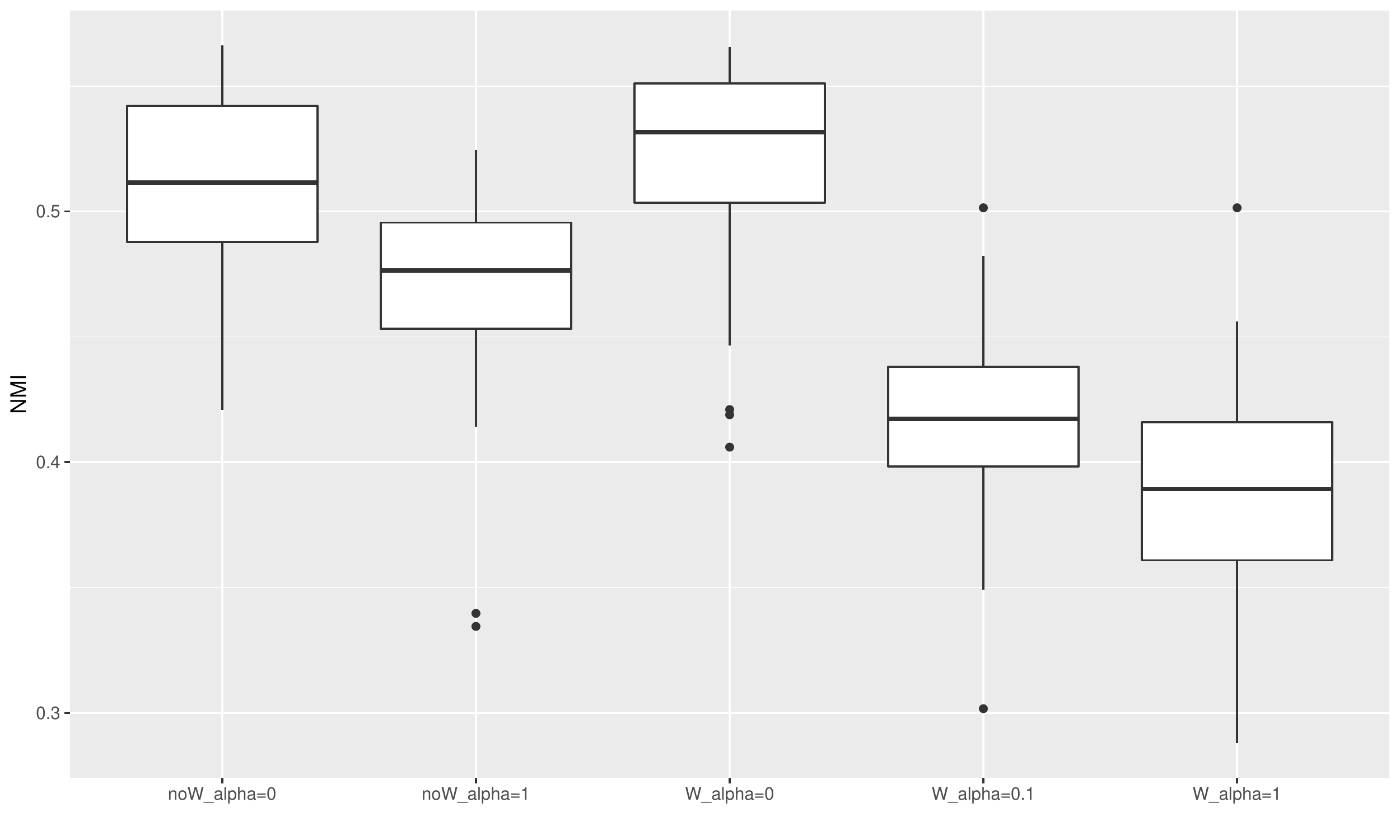}
\end{figure*}
For case I (both separate well), from Figure \ref{fig:figure 7} we can see that all five models are not bad in clustering and parameter estimation, this makes sense because both network and features separate well in case I. However, the mixture model with weights are better in parameter estimation, especially in the estimation of latent class proportions $\lambda$ and network connections $\phi$. The model with our choosen $\alpha = 0.1$ (W-alpha=0) has smaller number of mis-clusterings than the model with $\alpha=1$ (W-alpha=1) because it has larger NMI and similar modularity, as we can see from the bottom two plots. The model noW-alpha=1 has the smallest number of mis-clusterings, because it has similar NMI and larger modularity as we can see from Figure \ref{fig:figure 3}. Therefore, modularity and NMI reflect clustering quality. We can use them to get some idea about clustering even though we don't have true labels in real data.\\
\begin{figure*}
    \centering
    \caption{Parameter estimations, Number of mis-clusterings, Modularity and NMI by using different models for case II (features separate well) when n=300}
    \label{fig:figure 8}
    \includegraphics[scale=0.4]{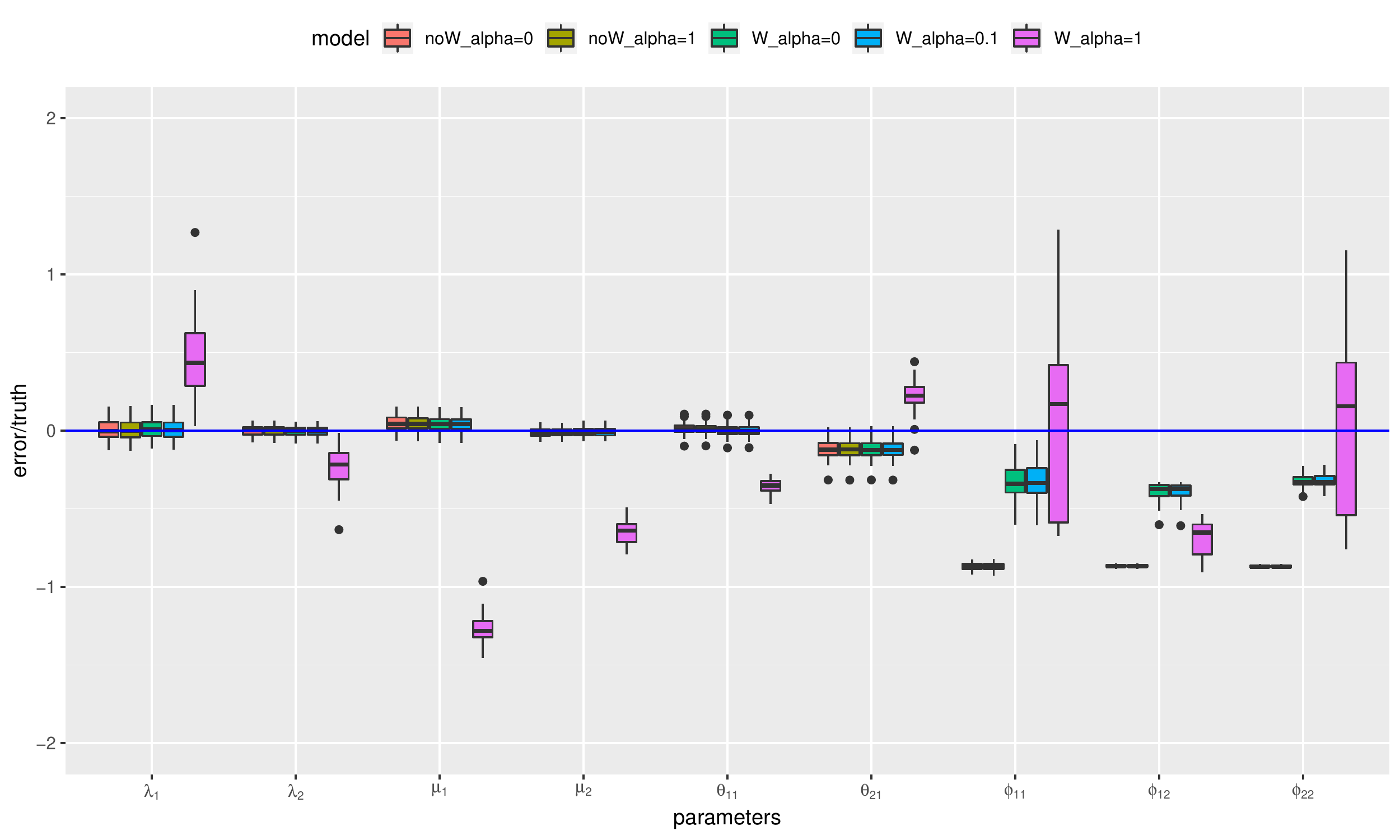}
    \includegraphics[scale=0.4]{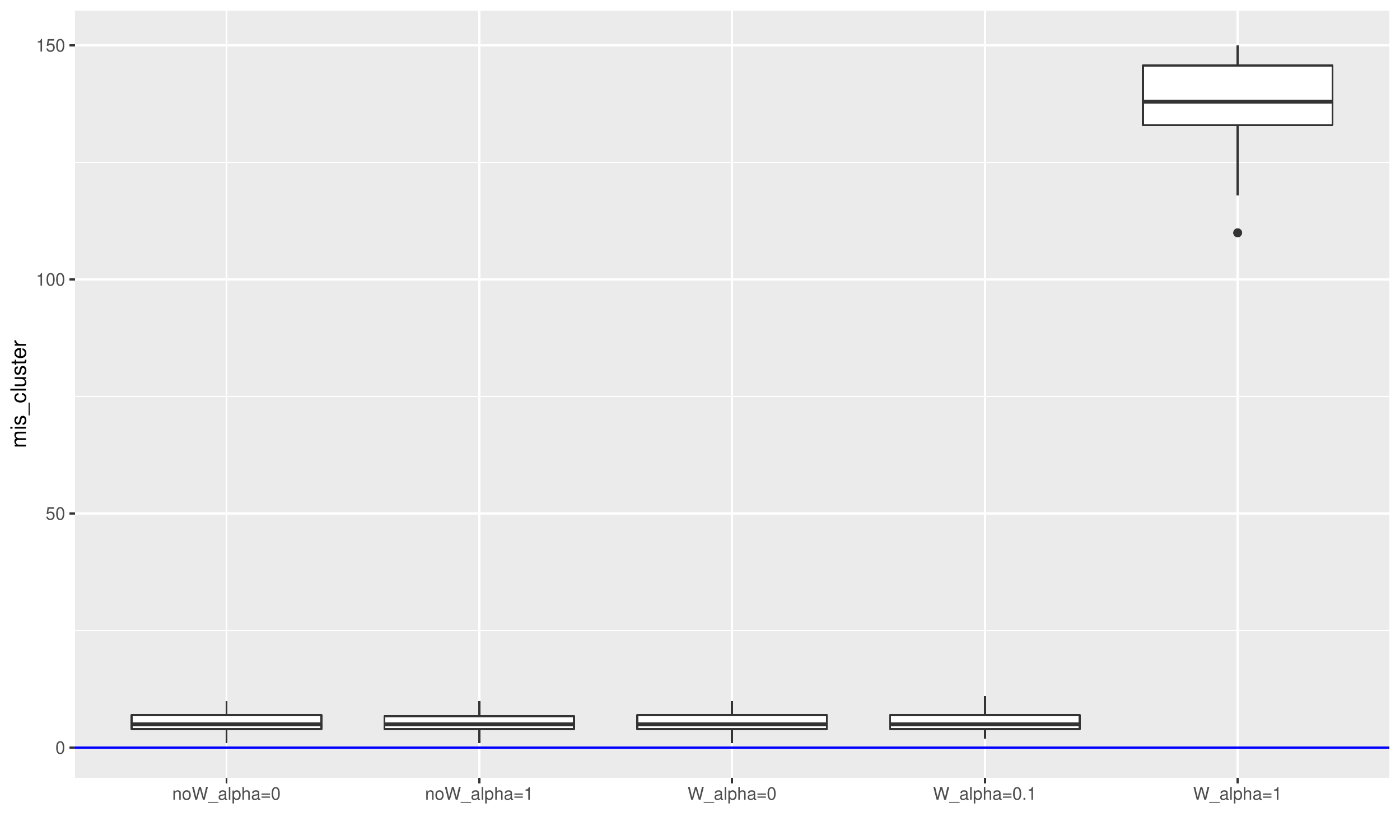}
    \includegraphics[scale=0.24]{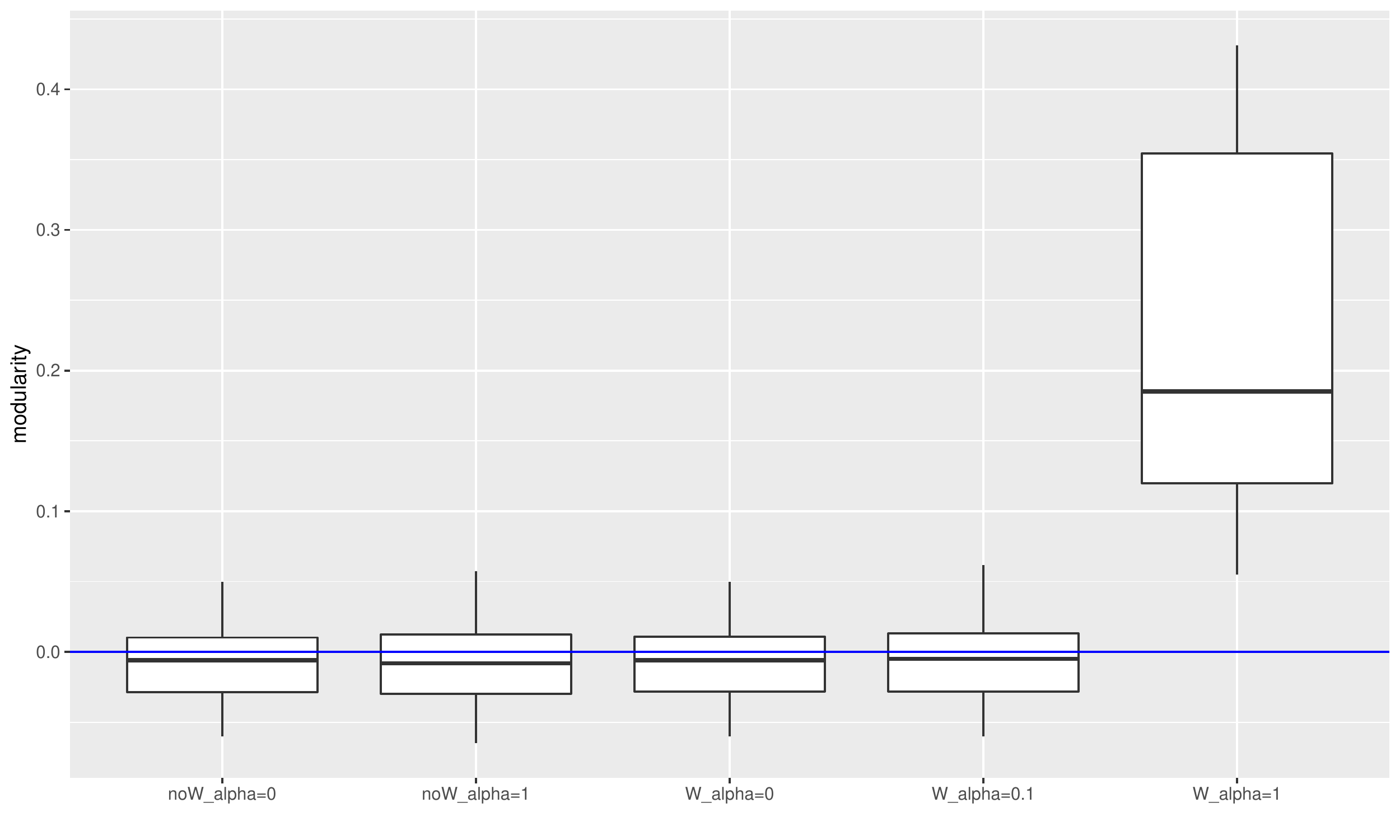}
    \includegraphics[scale=0.24]{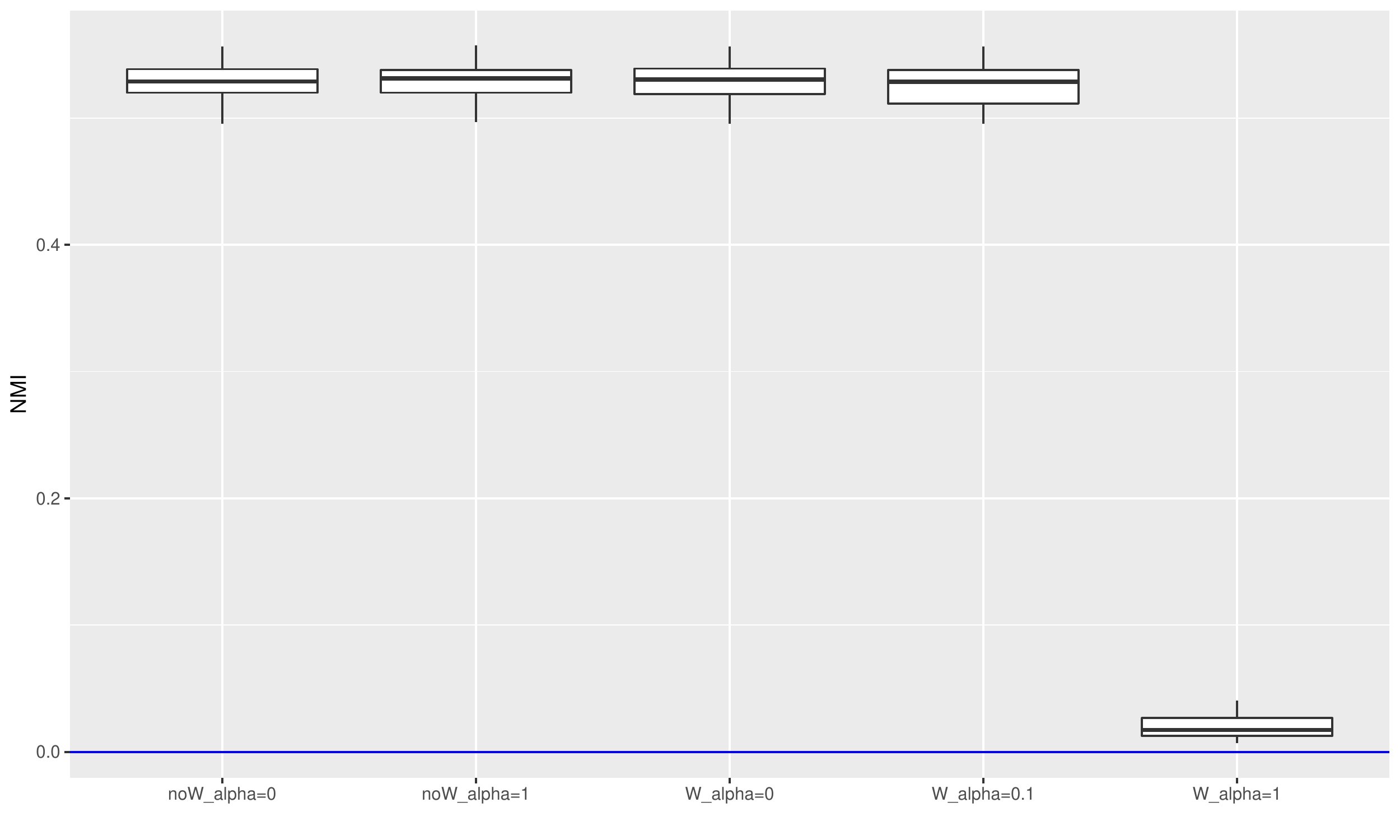}
\end{figure*}
For case II (only features separate well), Figure \ref{fig:figure 8} shows that when node features separate well, but the network does not, all models except the weighted model with $\alpha=1$, get pretty good clustering results. Also, the model with weighting gives better network structure parameter estimation $\phi$. This case tells us that when only node features are important and have obvious communities, the tuning parameter is essential to avoid overfitting of the noisy network structure.\\
\begin{figure*}
    \centering
    \caption{Parameter estimations, Number of mis-clusterings, Modularity and NMI by using different models for case III (network separate well) when n=300}
    \label{fig:figure 9}
    \includegraphics[scale=0.4]{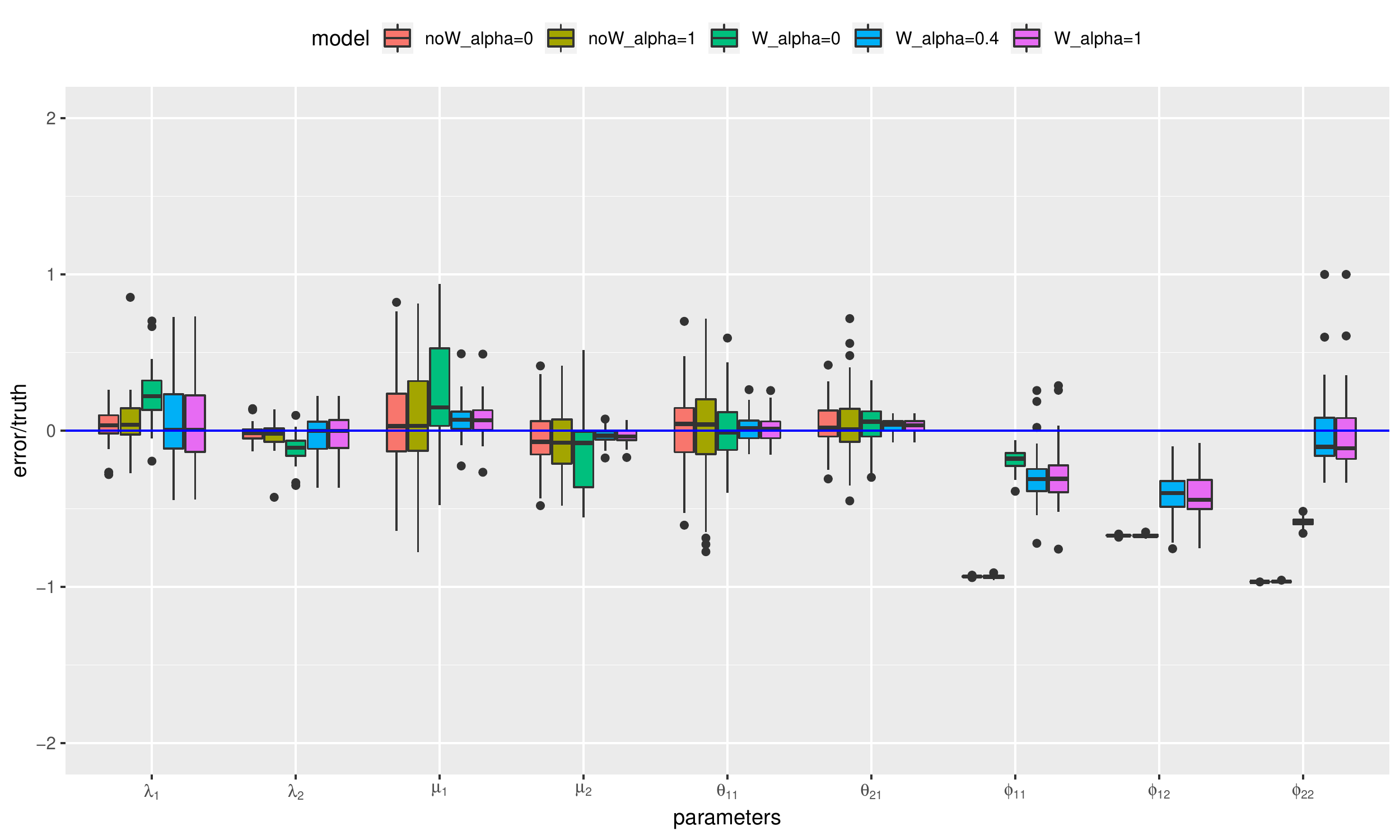}
    \includegraphics[scale=0.4]{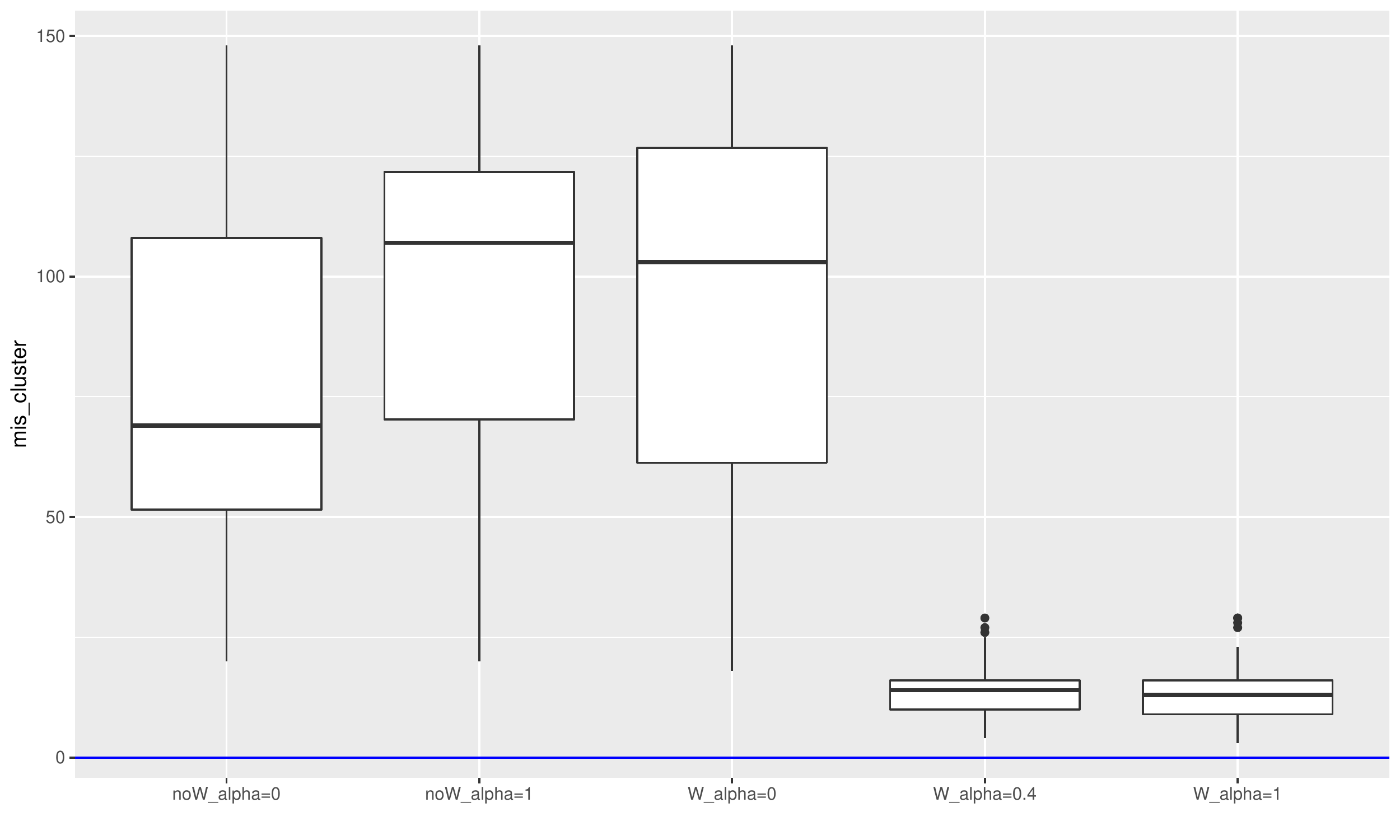}
    \includegraphics[scale=0.24]{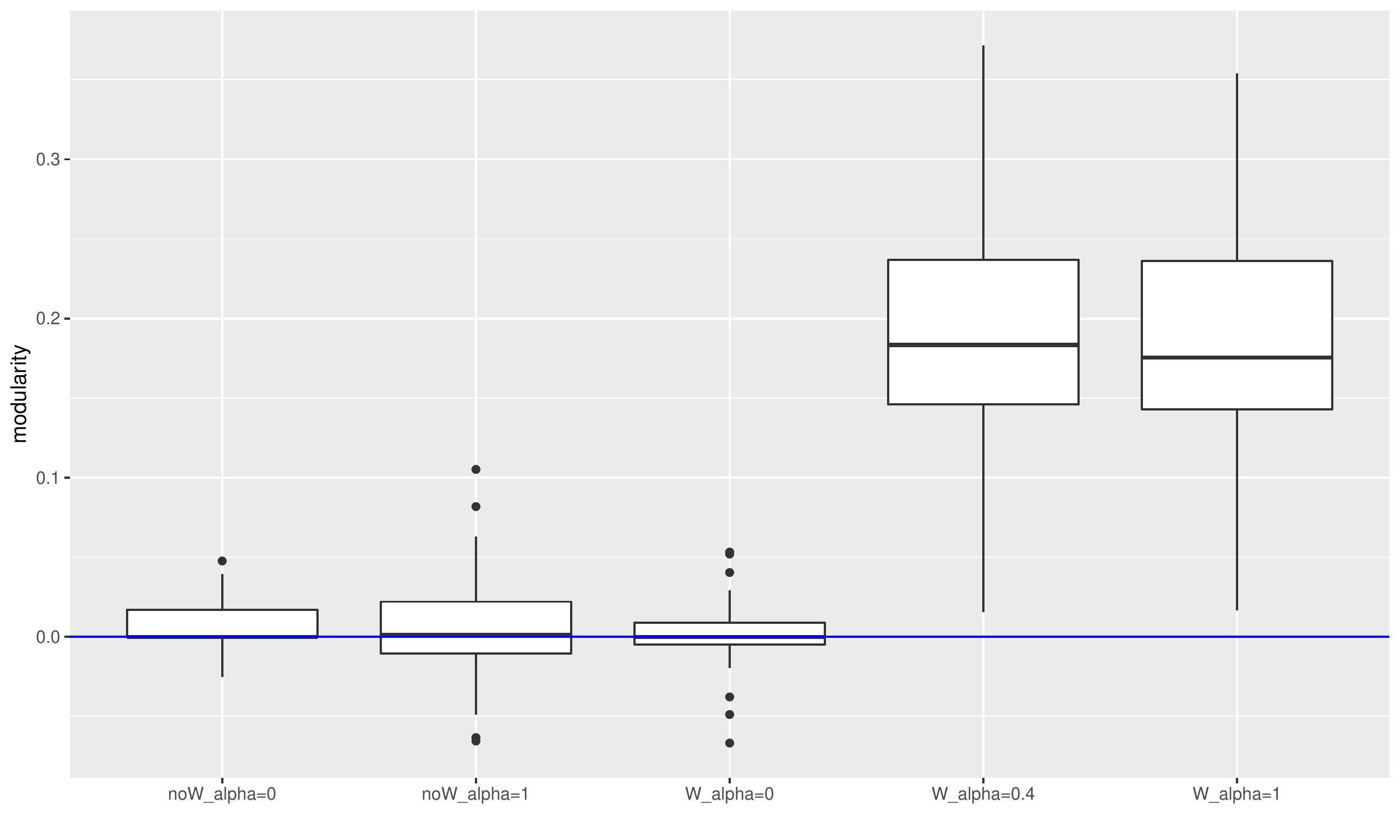}
    \includegraphics[scale=0.24]{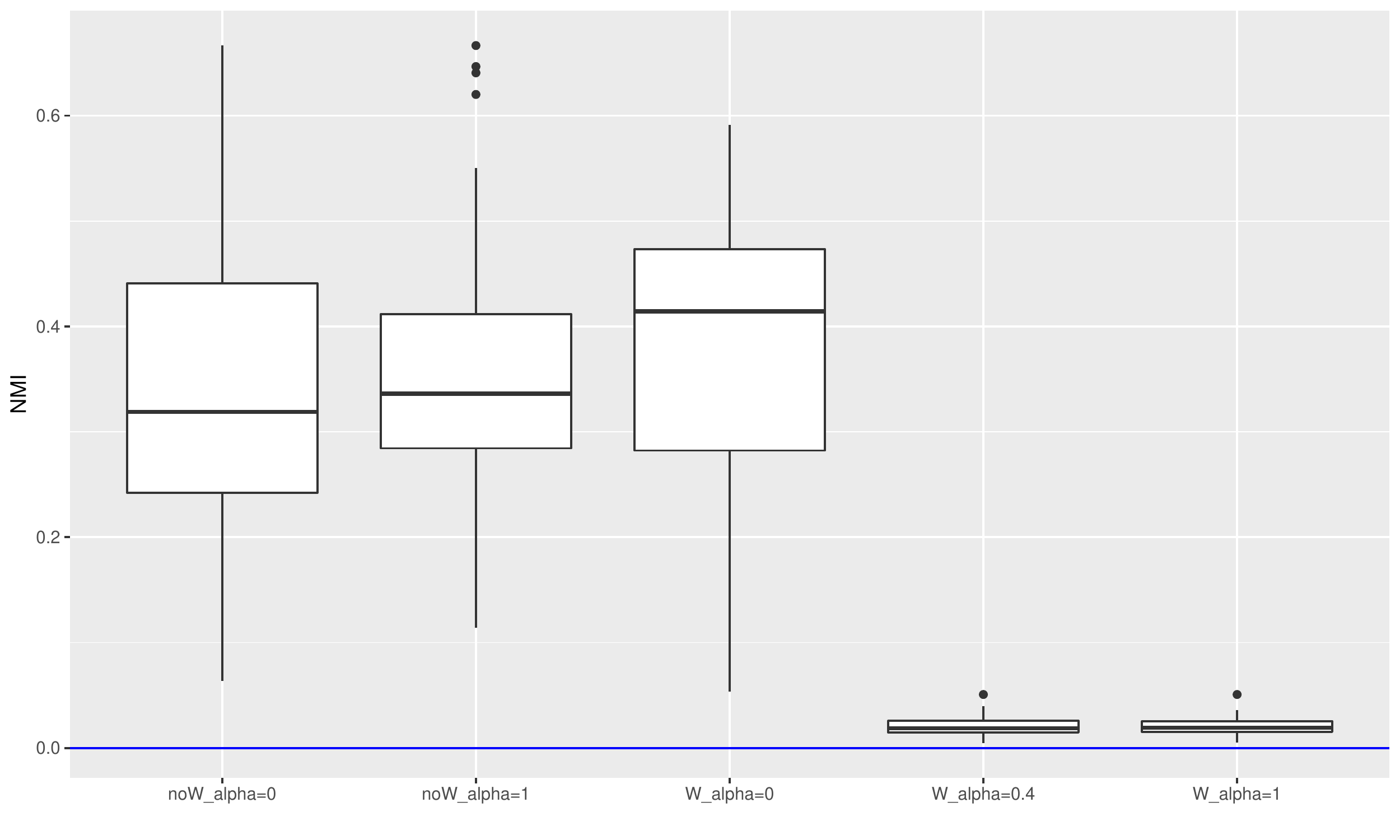}
\end{figure*}
Figure \ref{fig:figure 9} are the result for the third case, only the network structure separates well. We can still see models with weighting give better parameter estimates. In this case, we can also see that the models with weights and larger tuning parameter (W-alpha=0.4 and W-alpha=1) have better clustering results. This case tells us that using a larger tuning parameter is important when the network has clear communities. Both case II and case III support the importance of the tuning parameter in clustering sampled network data with node features by using the weighted mixture model. \\
\begin{figure*}
    \centering
    \caption{Parameter estimations, Number of mis-clusterings, Modularity and NMI by using different models for case IV (both do not separate well) when n=300}
    \label{fig:figure 10}
    \includegraphics[scale=0.4]{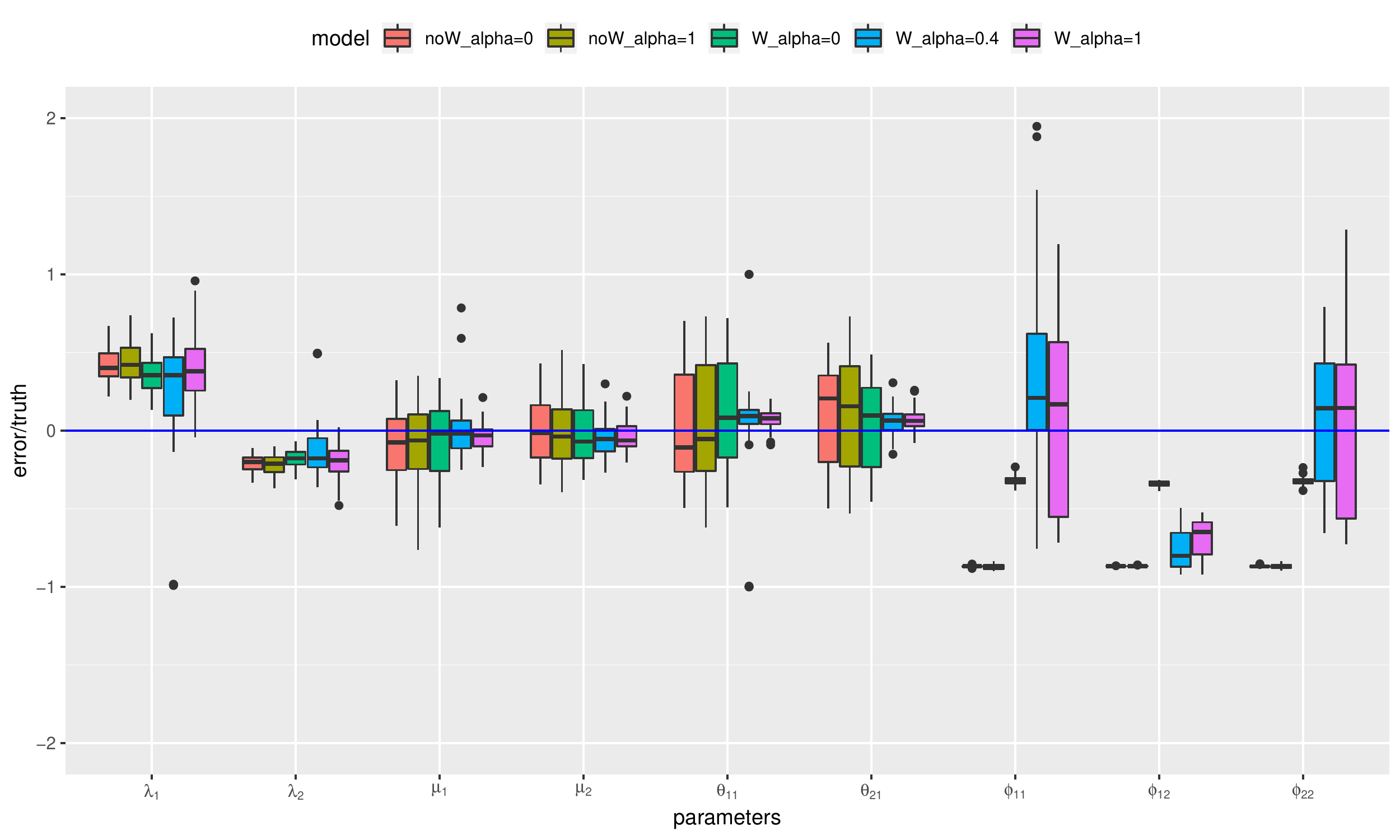}
    \includegraphics[scale=0.4]{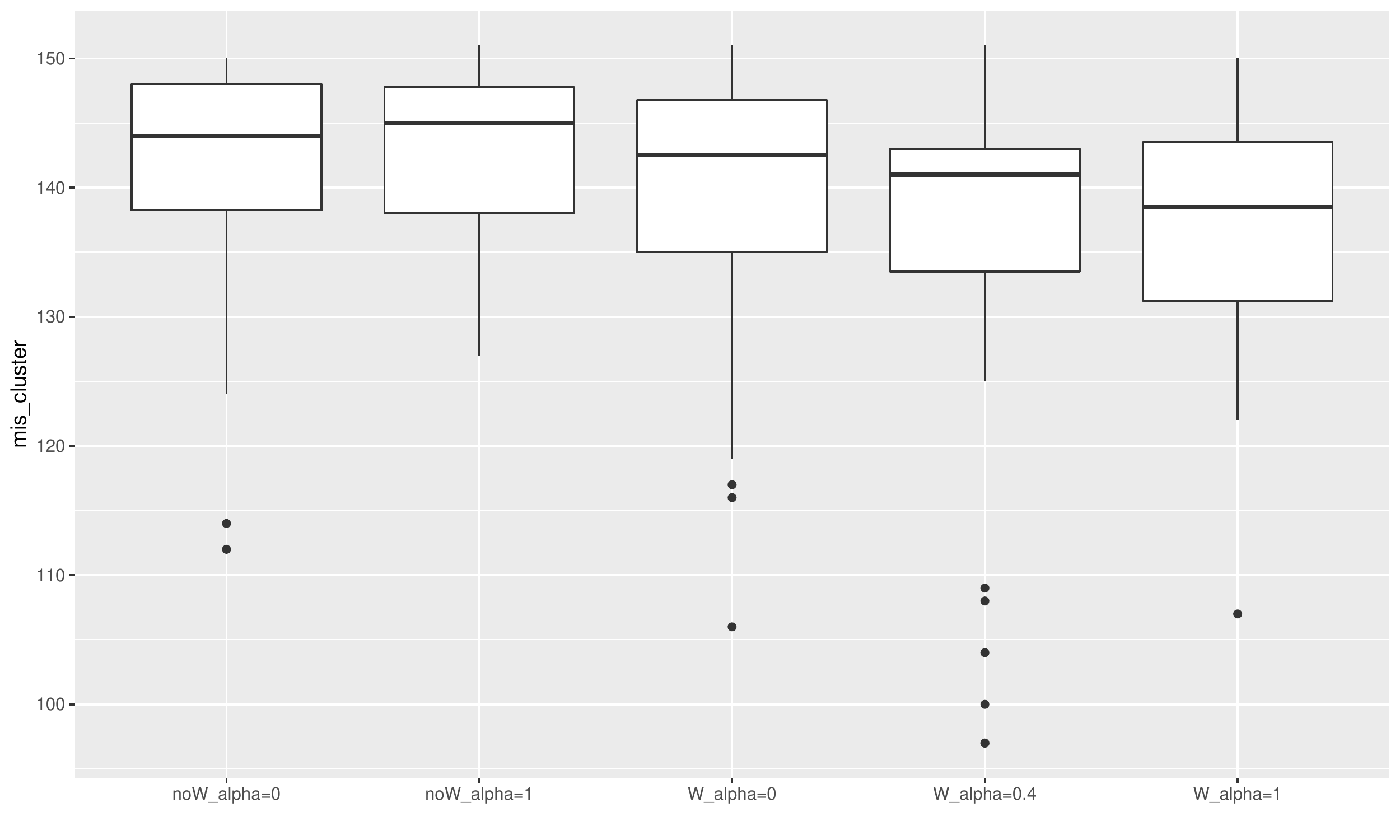}
    \includegraphics[scale=0.24]{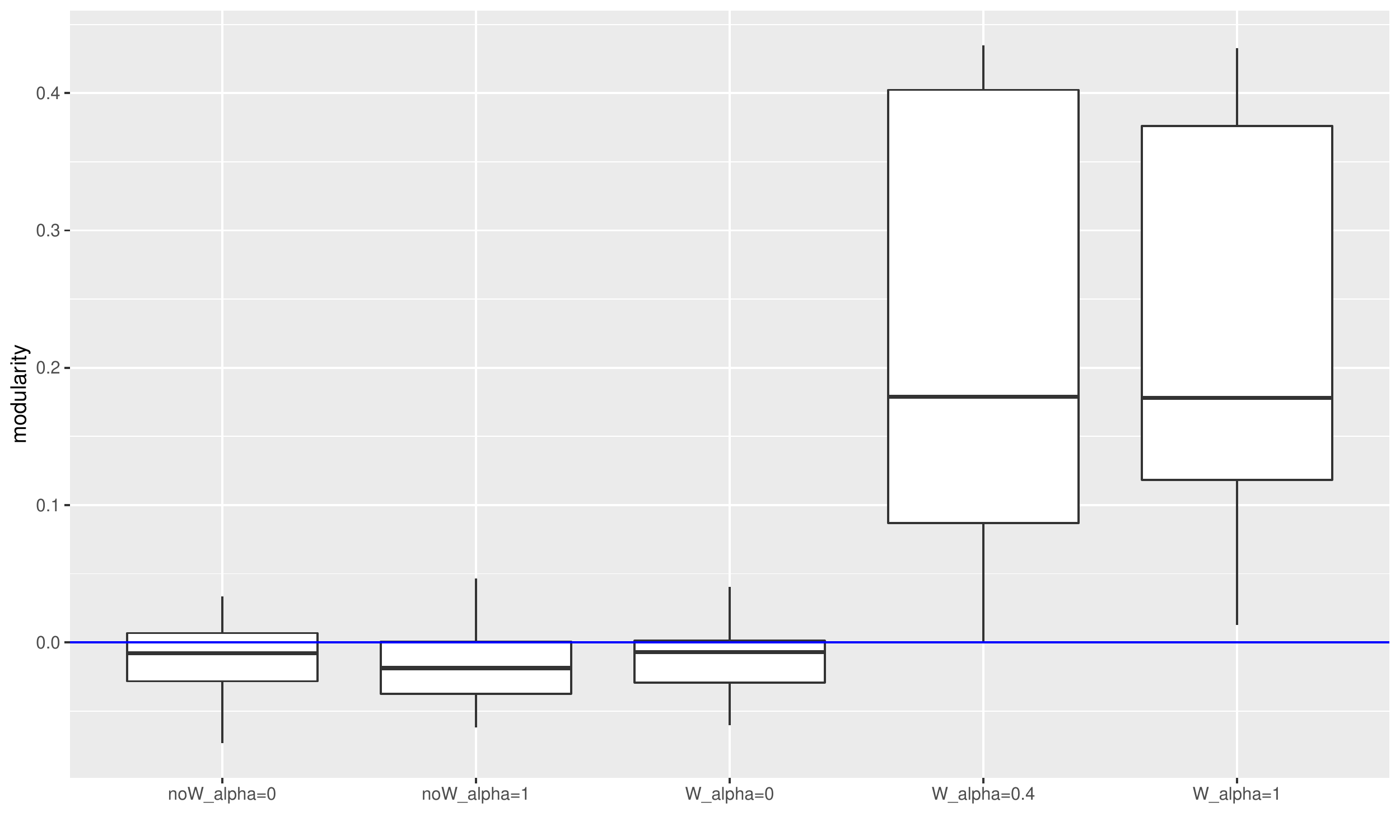}
    \includegraphics[scale=0.24]{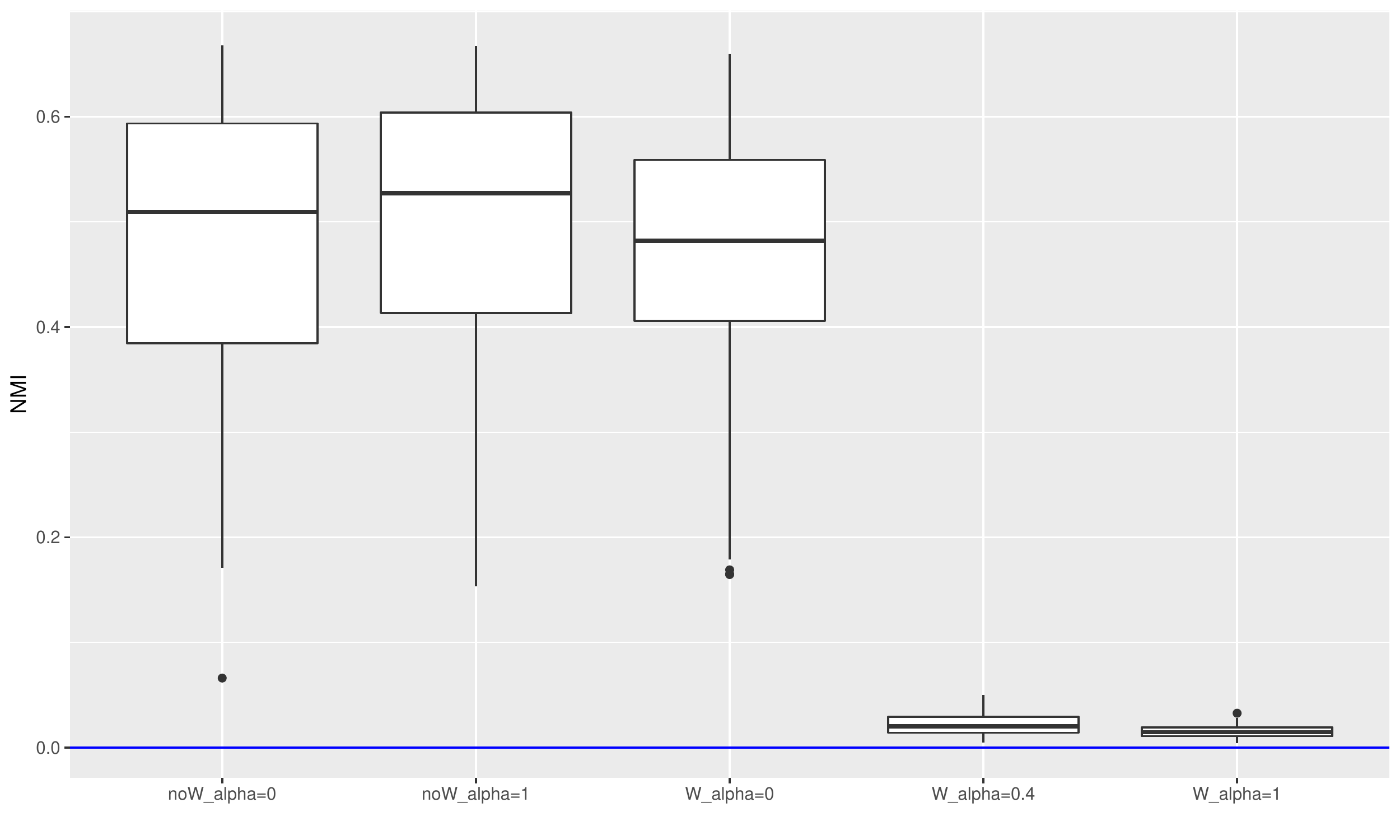}
\end{figure*}
For the last case, when both network and features don't separate well, Figure \ref{fig:figure 10} shows that all methods give large numbers of mis-clusterings. But we can see that models with weighting still give better parameter estimates.\\
\begin{table*}
    \centering
    \caption{Parameter summary statistics, mean (the first number in each cell), standard deviation (the second number) and MSE (the third number), under different methods when $n=300$ and $n=100$ for case I; M1 (noW-alpha=0), M2 (noW-alpha=1), M3 (noW-alpha-star), M4 (W-alpha=0), M5 (W-alpha=1), M6 (W-alpha-star)}
    \label{tab:table_MSE}
    \smallskip\noindent
    \scalebox{0.7}{
    \begin{tabular}{|c|c|c|c|c|c|c|c|c|c|c|c|c|} %
    \hline
    \multirow{2}{*}{Case I} & \multicolumn{6}{c|}{n=300} &\multicolumn{6}{c|}{n=100}\\
    &M1 &M2 &M3 &M4 &M5 &M6 &M1 &M2 &M3 &M4 &M5 &M6\\
    \hline
         $\lambda_1(=0.33)$ &\makecell{0.24\\(0.06)\\(0.01)} &\makecell{0.17\\(0.06)\\(0.03)} &\makecell{0.17\\(0.06)\\(0.02)} &\makecell{0.34\\(0.08)\\(0.006)} &\makecell{0.34\\(0.09)\\(0.007)} &\makecell{0.34\\(0.09)\\(0.008)} &\makecell{0.18\\(0.1)\\(0.03)} &\makecell{0.16\\(0.09)\\(0.04)} &\makecell{0.16\\(0.09)\\(0.04)} &\makecell{0.31\\(0.14)\\(0.02)} &\makecell{0.33\\(0.15)\\(0.02)} &\makecell{0.33\\(0.15)\\(0.02)} \\
         \hline
         $\lambda_2(=0.67)$ &\makecell{0.76\\(0.06)\\(0.01)} &\makecell{0.83\\(0.06)\\(0.03)} &\makecell{0.83\\(0.06)\\(0.02)} &\makecell{0.66\\(0.08)\\(0.006)} &\makecell{0.66\\(0.09)\\(0.007)} &\makecell{0.66\\(0.09)\\(0.008)} &\makecell{0.82\\(0.1)\\(0.03)} &\makecell{0.84\\(0.09)\\(0.04)} &\makecell{0.84\\(0.09)\\(0.04)}
         &\makecell{0.69\\(0.14)\\(0.02)} &\makecell{0.67\\(0.15)\\(0.02)} &\makecell{0.67\\(0.15)\\(0.02)}\\
         \hline
         $\mu_1(=-2)$ &\makecell{-1.19\\(0.8)\\(1.38)} &\makecell{-1.87\\(0.3)\\(0.09)} &\makecell{-1.87\\(0.3)\\(0.5)} &\makecell{-1.69\\(0.5)\\(0.3)} &\makecell{-1.54\\(0.2)\\(0.3)} &\makecell{-1.60\\(0.2)\\(0.2)} &\makecell{-1.44\\(1.1)\\(1.53)} &\makecell{-1.78\\(0.9)\\(0.88)} &\makecell{-1.79\\(0.9)\\(1.12)} &\makecell{-1.88\\(0.7)\\(0.46)} &\makecell{-1.47\\(0.7)\\(0.71)} &\makecell{-1.51\\(0.6)\\(0.54)} \\
         \hline
         $\mu_2(=2)$ &\makecell{2.02\\(0.11)\\(0.01)} &\makecell{1.95\\(0.06)\\(0.006)} &\makecell{1.95\\(0.06)\\(0.007)} &\makecell{1.99\\(0.1)\\(0.01)} &\makecell{1.95\\(0.06)\\(0.005)} &\makecell{1.97\\(0.05)\\(0.004)} &\makecell{2\\(0.12)\\(0.02)} 
         &\makecell{2\\(0.13)\\(0.02)} 
         &\makecell{2\\(0.13)\\(0.01)} 
         &\makecell{1.99\\(0.13)\\(0.02)} &\makecell{1.95\\(0.13)\\(0.02)} &\makecell{1.96\\(0.13)\\(0.02)} \\
         \hline
         $\theta_1(=0.8)$ &\makecell{0.76\\(0.07)\\(0.007)} &\makecell{0.78\\(0.06)\\(0.004)} &\makecell{0.78\\(0.06)\\(0.005)} &\makecell{0.77\\(0.06)\\(0.005)} &\makecell{0.74\\(0.05)\\(0.007)} &\makecell{0.75\\(0.06)\\(0.006)} &\makecell{0.78\\(0.16)\\(0.03)} &\makecell{0.78\\(0.16)\\(0.02)} &\makecell{0.78\\(0.16)\\(0.03)} &\makecell{0.8\\(0.14)\\(0.02)} &\makecell{0.73\\(0.17)\\(0.03)} &\makecell{0.73\\(0.17)\\(0.03)} \\
         \hline
         $\theta_2(=0.4)$ &\makecell{0.35\\(0.04)\\(0.004)} &\makecell{0.37\\(0.02)\\(0.001)} &\makecell{0.37\\(0.02)\\(0.002)}
         &\makecell{0.36\\(0.03)\\(0.003)} &\makecell{0.37\\(0.02)\\(0.001)} &\makecell{0.37\\(0.02)\\(0.001)} &\makecell{0.38\\(0.06)\\(0.004)} &\makecell{0.39\\(0.06)\\(0.004)} &\makecell{0.39\\(0.06)\\(0.004)} &\makecell{0.39\\(0.06)\\(0.004)} &\makecell{0.4\\(0.06)\\(0.003)} 
         &\makecell{0.4\\(0.06)\\(0.003)} \\
         \hline
         $\phi_{11}(=0.1)$ &\makecell{0.015\\(0.006)\\(0.007)} &\makecell{0.027\\(0.007)\\(0.005)} &\makecell{0.027\\(0.006)\\(0.006)} &\makecell{0.05\\(0.008)\\(0.002)} &\makecell{0.07\\(0.016)\\(0.001)} &\makecell{0.07\\(0.015)\\(0.001)} &\makecell{0.06\\(0.05)\\(0.004)} &\makecell{0.09\\(0.07)\\(0.004)} &\makecell{0.09\\(0.06)\\(0.004)} &\makecell{0.05\\(0.03)\\(0.003)} &\makecell{0.09\\(0.06)\\(0.004)} &\makecell{0.09\\(0.06)\\(0.004)} \\
         \hline
         $\phi_{12}(=0.02)$ &\makecell{0.004\\(0.001)\\(3e-4)} &\makecell{0.002\\(0.0005)\\(3e-4)} &\makecell{0.002\\(0.0006)\\(3e-4)} &\makecell{0.03\\(0.011)\\(2e-4)} &\makecell{0.01\\(0.003)\\(8e-5)} &\makecell{0.01\\(0.003)\\(9e-5)} &\makecell{0.01\\(0.004)\\(8e-5)} &\makecell{0.009\\(0.004)\\(1e-4)} &\makecell{0.009\\(0.004)\\(1e-4)} &\makecell{0.02\\(0.01)\\(1e-4)} &\makecell{0.01\\(0.006)\\(8e-5)} &\makecell{0.01\\(0.006)\\(1e-4)} \\
         \hline
         $\phi_{22}(=0.2)$ &\makecell{0.007\\(0.0003)\\(0.04)} &\makecell{0.008\\(0.0004)\\(0.04)} &\makecell{0.008\\(0.0004)\\(0.04)} &\makecell{0.16\\(0.02)\\(0.002)} &\makecell{0.19\\(0.04)\\(0.002)} &\makecell{0.19\\(0.04)\\(0.002)} &\makecell{0.02\\(0.001)\\(0.03)} &\makecell{0.02\\(0.002)\\(0.03)} &\makecell{0.02\\(0.002)\\(0.03)} &\makecell{0.15\\(0.03)\\(0.004)} &\makecell{0.18\\(0.05)\\(0.004)} &\makecell{0.18\\(0.06)\\(0.004)} \\
         \hline
    \end{tabular}
    }
\end{table*}
To study the effect of sample size, we do the same work for the RDS sample data with $n=100$ and the results are summarized in Table \ref{tab:table_MSE}. We can see that mixture models with weights still give less biased parameter estimates, but the uncertainty of parameter estimates are larger when $n=100$.\\ 
From all the simulation result we find that the mixture model with weights gives better parameter estimates. Adding tuning parameter $\alpha$ is essential in finding more interpretable communities. Modularity and normalized mutual information help to determine reasonable tuning parameter values and give us information about the quality of the clustering result. 
\section{Application} \label{Application}
In this section, we apply the mixture models with and without weights to cluster RDS data collected on young adult opioid users in New York City (NYC).\\
\textbf{Young adult opioid users RDS data in NYC}\\
The data we use are RDS data sampled from opioid users aged 18-29 who had non-medical use of prescription opioids and/or heroin in the past 30 days, currently living in NYC, speak English and are able to provide informed consent. Each participant was interviewed for personal demographic information and drug use behavioral questions. Since participants in this network are recruited through referral, it is believed that community structure exists in this observed recruitment network. To detect those communities, we apply the weighted log-likelihood mixture model with choosen tuning parameter to the NYC young adult opioid users data. Node features used for this clustering are age, borough, opioid injection years, other drugs injection years, homeless, how many are older than 29 among people you know that use POs and live in NYC (NetChar4) and how many inject drugs among people you know that use opioids and live in NYC (NetChar22). The clustering results are summarized in Tables \ref{tab: table 5} and \ref{tab: table 6} and Figure \ref{fig: figure 2}.\\
\begin{figure*}
    \centering
    \caption{Modularity and NMI vs $\alpha$ in the weighted and un-weighted mixture model for the Opioid users RDS data}
    \label{fig:figure_nyc}
    \includegraphics[scale=0.8]{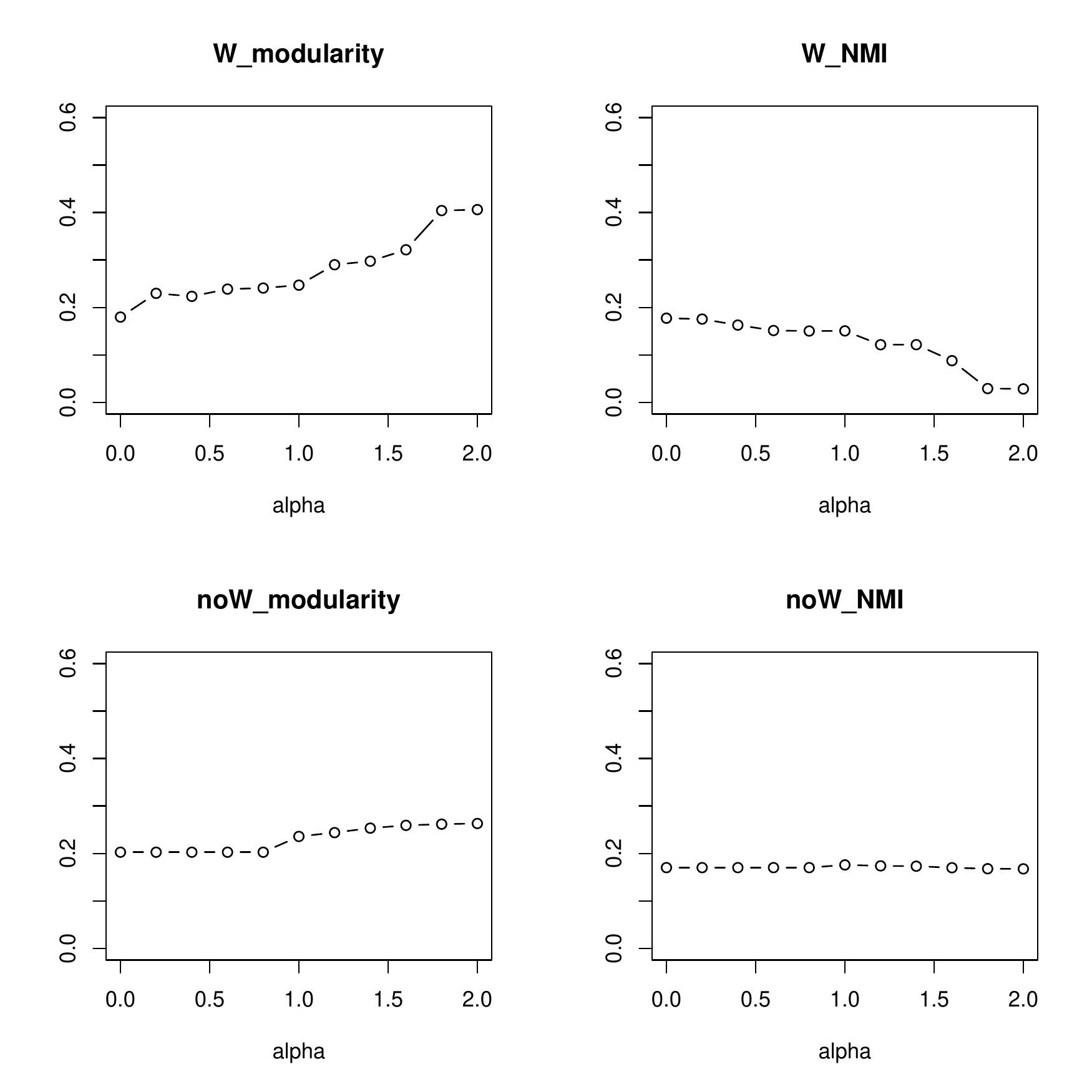}
\end{figure*}
To balance opioid users' attributes and their network connections, we first find a tuning parameter. From Figure \ref{fig:figure_nyc} we can see that the modularity is not small for this sampled network dataset which indicates social communities exist in the opioid users' RDS dataset. When $\alpha=0$, modularity and NMI are around 0.2. We conclude that communities based on node features explains some community structures of the network which is reasonable for our opioid users RDS dataset because opioid users with similar use behavior are more likely to be connected. \\
In the mixture model with weighting, the modularity increases and NMI decreases. We choose $\alpha = 1$ as our tuning parameter values because the corresponding NMI values are still not very small and the modularity values are relatively large. In this way, our clustering result is based on both node features and network structure. For the model without weight, $\alpha = 1$ is also reasonable because NMI does not change much with different $\alpha$ values but modularity increase more appreciably from 0.8 to 1.0.\\
\begin{figure*}
    \caption{Clustering result using mixture model with and without weights on young adult opioid users RDS data in NYC}
    \label{fig: figure 2}
    \includegraphics[scale=0.4]{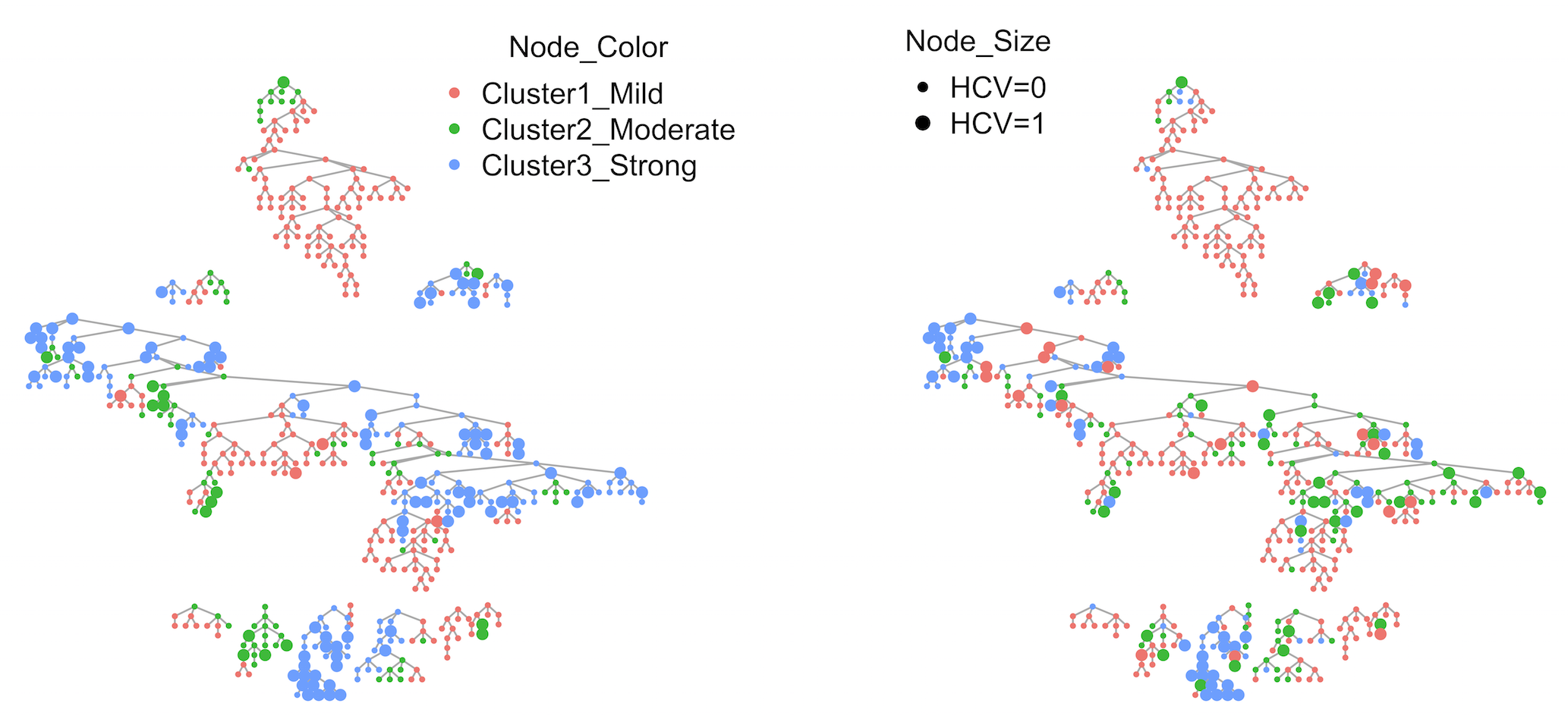}
\end{figure*}
\begin{table*}
    \centering
    \caption{Feature Comparisons based on clustering from weighted log-likelihood mixture model with $\alpha=1$ on the young adults opioid users RDS data in NYC.}
     \label{tab: table 5}
    \smallskip\noindent
    \scalebox{0.9}{
    \begin{tabular}{|c|c|c|p{1cm}|p{1cm}|p{1.5cm}|p{2.5cm}|p{2.5cm}|p{1.5cm}|}
    \hline
     Cluster &Prop & Prop-HCV &Age &Inj-years &Inj-Others-years &Prop-(NetChar4$\geq$5)&Prop-(NetChar22$\geq$5)&Prop-Homeless\\
     \hline 
     Strong & 0.36 &0.43 & 25 &5 &6.6 & 0.4 &0.74 &0.5\\
     \hline
     Moderate & 0.21 &0.17 & 24.5 &3.9 &2.1 & 0.27 &0.66 &0.17\\
     \hline
     Mild & 0.43 &0.016 & 23 &0 &0.6 & 0.2 &0.21 &0.09\\
     \hline
    \end{tabular}
    }
    \begin{itemize}
    \item Prop: proportion of sample in each cluster.
    \item Prop-HCV: proportion of HCV position.
    \item Age, Inj-years, Inj-Others-year: average age, opioid injection years and others drugs injection years in each cluster.
    \item NetChar4: how many are older than 29 among people you know that use opioids and live in NYC?
    \item NetChar22: how many inject drugs among people you know that use opioids and live in NYC?
    \item Prop-(NetChar4$\geq 5$): sample proportion in each cluster with NetChar4 $\geq 5$.
    \item Prop-(NetChar22$\geq 5$): sample proportion in each cluster with NetChar22 $\geq 5$.
    \item Prop-Homeless: proportion of homeless people in each cluster.
\end{itemize}
\end{table*}
Hepatitis C Virus (HCV) is not included in the clustering model, but from the clustering result graph Figure \ref{fig: figure 2}, we can see that the weighted mixture model is more likely to group people with HCV in cluster 1, which contains most heavy opioid drug users. 43.4\% people in cluster 1 are HCV positive based on Table \ref{tab: table 5}. Also, based on Table \ref{tab: table 5}, cluster 1 has people with larger age values, more opioid and drug injectors, people who know more opioid users older than 29 and know more drug injectors, and much more homeless than cluster 2 and cluster 3. Cluster 2 contains moderately risky opioid users. Although average age in it is similar to average age in cluster 1, people in cluster 2 are much newer in terms of injection years, they know fewer 29+ years old opioid users and most of them are not homeless. Cluster 3 is the least risky opioid users group because most of them are young, do not inject, know many fewer older opioid users and injectors. Overall, these three clusters separate opioid drug users very well in terms of those characteristics and drug use behaviors. \\
Table \ref{tab: table 6} tells us that participants from Bronx and Brooklyn are more likely in the mild cluster (cluster 3), samples from Queens and State Island are more likely to be in the strong cluster (cluster 1). Participants from Manhattan are evenly clustered into strong and mild groups, which we can see from Figure \ref{fig: figure 2} that the tree on the top has most of its samples coming from Manhattan and most of them are not homeless. Other people from Manhattan in other trees have much more homelessness. This supports the clustering result that about half participants from Manhattan are mild opioid drug users and half are strong opioid drug users. \\
\begin{table*}
    \centering
    \caption{Sample proportion by clusters from weighted log-likelihood mixture model in each borough for the young adults opioid users' RDS data in NYC.}
    \label{tab: table 6}
    \smallskip\noindent
    \scalebox{0.8}{
    \begin{tabular}{|c|c|c|p{2cm}|p{2cm}|p{2cm}|p{2cm}|p{2cm}|}
    \hline
     Cluster &Count &Prop &Prop-Manhattan &Prop-State Island &Prop-Brooklyn &Prop-Bronx &Prop-Queens\\
     \hline 
     Strong & 192 & 0.36 &0.47 &0.49 &0.24 & 0.17 &0.29\\
     \hline
     Moderate & 110 & 0.21 & 0.04 &0.3 &0.26 & 0.17 &0.51\\
     \hline
     Mild & 230 & 0.43 & 0.48 &0.21 &0.49 & 0.67 &0.20\\
     \hline
    \end{tabular}
    }
\end{table*}
With estimated network connection parameter $\hat{\phi}$, we can clearly see that people in the same cluster have more ties than people from different clusters. Among connections between two different clusters, people from moderate and mild clusters have much stronger cross-cluster connections than people from strong and moderate clusters, people between strong and mild clusters are least likely to be connected. This tells us that mild opioid drug users are much more likely to be influenced by moderate opioid drug users than strong opioid drug users, which targets the population we should focus on for intervention to protect young mild and potential opioid drug users. \\
Meanwhile, we also applied the mixture model without weights to cluster this NYC young adults opioid users' RDS data. Its clustering result is included in Figure \ref{fig: figure 2}. From Figure \ref{fig: figure 2} we can see that the weighted log-likelihood mixture model clusters more people in the strong opioid drug user group (cluster 1). This is because the weighted log-likelihood mixture model detects network structure better than the one without weights, which results in a clearer social connection effect in the clustering result. Capturing social connection effect is important in the NYC young adults opioid users' RDS data because it gives us guidelines for future interventions. \\
The network connection parameter estimation (assumed the full network size $N=1e4$) based on the weighted log-likelihood mixture model with $\alpha=1$ is 
 \[\hat{\phi} = 
\begin{blockarray}{cccc}
Strong & Moderate &Mild \\
\begin{block}{[ccc]c}
 0.015& 0.0005 &0.0002 &Strong \\
 0.0005 & 0.016 &0.001 &Moderate \\
 0.0002 & 0.0014 &0.009 &Mild\\
\end{block}
\end{blockarray}
\]
\section{Discussion and Conclusions}\label{Conclusion}
In this paper, we build a mixture model with weighted log-likelihood inference for clustering node-attributed RDS sample data. We also propose to add a tuning parameter to the weighted log-likelihood to balance contribution of node features and network structure in clustering. Node features in RDS network clustering enable us to understand how nodes differ across groups, and critically help to detect clusters despite the multiple isolated tree structures generated by the RDS. From the simulation study with two different RDS sample sizes, we see that the clustering algorithm is robust to the sample proportion. Adding weights as inverse sampling probabilities to the log-likelihood reduces bias in parameter estimation because RDS is not simple random sampling. Edge sampling probabilities are essential to capture the truth that two un-connected nodes in the RDS data does not necessarily mean they are not connected in the full network. This relates a very sparse RDS network to a less sparse underlying network. Weighted log-likelihood inference results in better network connection parameter estimation which tells us a closer truth about how strong the connections are within and between clusters in the underlying social network. To evaluate the clustering quality and find a proper tuning parameter value, we also discussed modularity and normalized mutual information and modified it for the pseudo-population network data. We recommend using these two metrics together to select a value for the tuning parameter. 
\pagebreak
\bibliography{aoas-RDS-Clustering}
\end{document}